\documentclass[journal,twoside,web]{ieeecolor}

\usepackage{generic} 
\usepackage{cite}
\usepackage{amsmath,amssymb,amsfonts}
\usepackage{algorithmic}
\usepackage{graphicx}
\usepackage{textcomp}
\usepackage{subcaption}
\let\labelindent\relax 
\usepackage{enumitem}

\usepackage{tikz}
\usetikzlibrary{shapes.geometric, arrows, backgrounds}

\tikzstyle{block} = [rectangle, rounded corners, minimum width=1.5cm, minimum height=1cm,text centered, draw=white, fill=blue!30]

\tikzstyle{blockspecial} = [rectangle, rounded corners, minimum width=1.5cm, minimum height=1cm,text centered, draw=white, fill=orange!30]

\tikzstyle{operation} = [circle, rounded corners, minimum width=0.5cm, minimum height=0.5cm,text centered, draw=black, fill=white!30]

\tikzstyle{mynode_state} = [circle, minimum size=1.5cm, text centered, draw=white, fill=orange!30] 

\tikzstyle{arrow} = [thick,->,>=stealth]

\def\BibTeX{{\rm B\kern-.05em{\sc i\kern-.025em b}\kern-.08em
    T\kern-.1667em\lower.7ex\hbox{E}\kern-.125emX}}
\begin{document}
\title{Numerical Gaussian process Kalman filtering for spatiotemporal systems}
\author{Armin K{\"u}per and Steffen Waldherr
\thanks{Armin K{\"u}per and Steffen Waldherr are with the CREaS group of the Chemical Engineering Department, KU Leuven, 3001 Leuven, Belgium (e-mail:$\{$armin.kuper, steffen.waldherr$\}$@kuleuven.be). }
}

\maketitle

\begin{abstract}
We present a novel Kalman filter for spatiotemporal systems called the numerical Gaussian process Kalman filter (GPKF).
Numerical Gaussian processes have recently been introduced as a physics informed machine learning method for simulating time-dependent partial differential equations without the need for spatial discretization.
We bring numerical GPs into probabilistic state space form.
This model is linear and its states are Gaussian distributed.
These properties enable us to embed the numerical GP state space model into the recursive Kalman filter algorithm.
We showcase the method using two case studies.
\end{abstract}


\section{Introduction}
\label{sec:intro}

State estimators are algorithms for reconstructing a system's state from a stream of noisy online measurements and model predictions.
The model is usually in the form of a probabilistic state space model.
Given linear finite-dimensional dynamics, the process and measurement equation read as 
\begin{align}
	\boldsymbol{x}_t &= \boldsymbol{A} \boldsymbol{x}_{t-1} + \boldsymbol{q}_{t-1}, \label{eq:processeqfinite} \\
	\boldsymbol{y}_t &= \boldsymbol{C} \boldsymbol{x}_t + \boldsymbol{r}_t,
	\label{eq:measeqfinite}
\end{align}
with the state at time $t$ as $\boldsymbol{x}_t \in \mathbb{R}^{d_x}$, and the measurements as $\boldsymbol{y}_t \in \mathbb{R}^{d_y}$.
Dynamics are described by matrix $\boldsymbol{A}$ while measurements are obtained through $\boldsymbol{C}$.
Process $\boldsymbol{q}_{t-1}$ and measurement noise $\boldsymbol{r}_t$ are modeled as Gaussian distributed zero-mean white noise with covariance matrix $\boldsymbol{Q}$ and $\boldsymbol{R}$, respectively.
Process and measurement equations \eqref{eq:processeqfinite}\,--\,\eqref{eq:measeqfinite} can also be represented by probability density functions $p(\cdot)$ that take the form of Gaussian distributions $\text{N}(\cdot)$ due to the nature of the noise processes.
In particular we have
\begin{align}
	p(\boldsymbol{x}_t | \boldsymbol{x}_{t-1}) &= \text{N} \left(\boldsymbol{x}_t  | \boldsymbol{A} \boldsymbol{x}_{t-1}, \boldsymbol{Q}   \right), \\
	p(\boldsymbol{y}_t | \boldsymbol{x}_t) &= \text{N} \left(\boldsymbol{y}_t | \boldsymbol{C} \boldsymbol{x}_t, \boldsymbol{R} \right).
\end{align}
The following question arises: How should the measurement $\boldsymbol{y}_t$ be used to correct the model prediction of $\boldsymbol{x}_t$ in real time?

One answer to this question is the Kalman filter.
In fact, it gives the optimal estimate $\boldsymbol{m}_t^{opt}$ of the current state $\boldsymbol{x}_t$ under consideration of the measurement history up to the present $\boldsymbol{y}_{1:t} = \{ \boldsymbol{y}_1,\hdots,\boldsymbol{y}_t \}$ \cite{kalman1960new}.
It is optimal in the sense that the expected squared error between estimated and true state conditioned on the measurement history is minimized.

The Kalman filter estimates the state in a recursive manner.
In Bayesian statistical terms this can be written as
\begin{align}
	p(\boldsymbol{x}_{t} | \boldsymbol{y}_{1:t}) &= \dfrac{p(\boldsymbol{y}_{t} | \boldsymbol{x}_{t}) p(\boldsymbol{x}_{t} | \boldsymbol{y}_{1:t-1})}{p(\boldsymbol{y}_{t} | \boldsymbol{y}_{1:t-1})}.
\end{align}
Since the recursion starts with a Gaussian distribution and the dynamics are linear, all the above distributions remain Gaussian, and the posterior estimate can be obtained in closed form by calculating its mean and covariance.

Remember that the above model equations described finite-dimensional systems.
Infinite-dimensional filtering theory certainly exists \cite{curtain2012introduction}, but direct practical infinite-dimensional Kalman filtering does not to the best knowledge of the authors. 
Finite-dimensional approximations are therefore needed.
Using spatial discretization methods, e.g. the finite volume method \cite{ferziger2002computational}, one usually obtains a large system of ordinary differential equations which can be formulated as the above state space model \eqref{eq:processeqfinite}\,--\,\eqref{eq:measeqfinite}.

\begin{figure}[tb!]
\begin{center}
  \resizebox{1\columnwidth}{!}{%
\begin{tikzpicture}[node distance=4.6cm]
	\node[inner sep=0pt] (base_plot) at (0,0) {
	    \includegraphics[width=0.6\columnwidth]{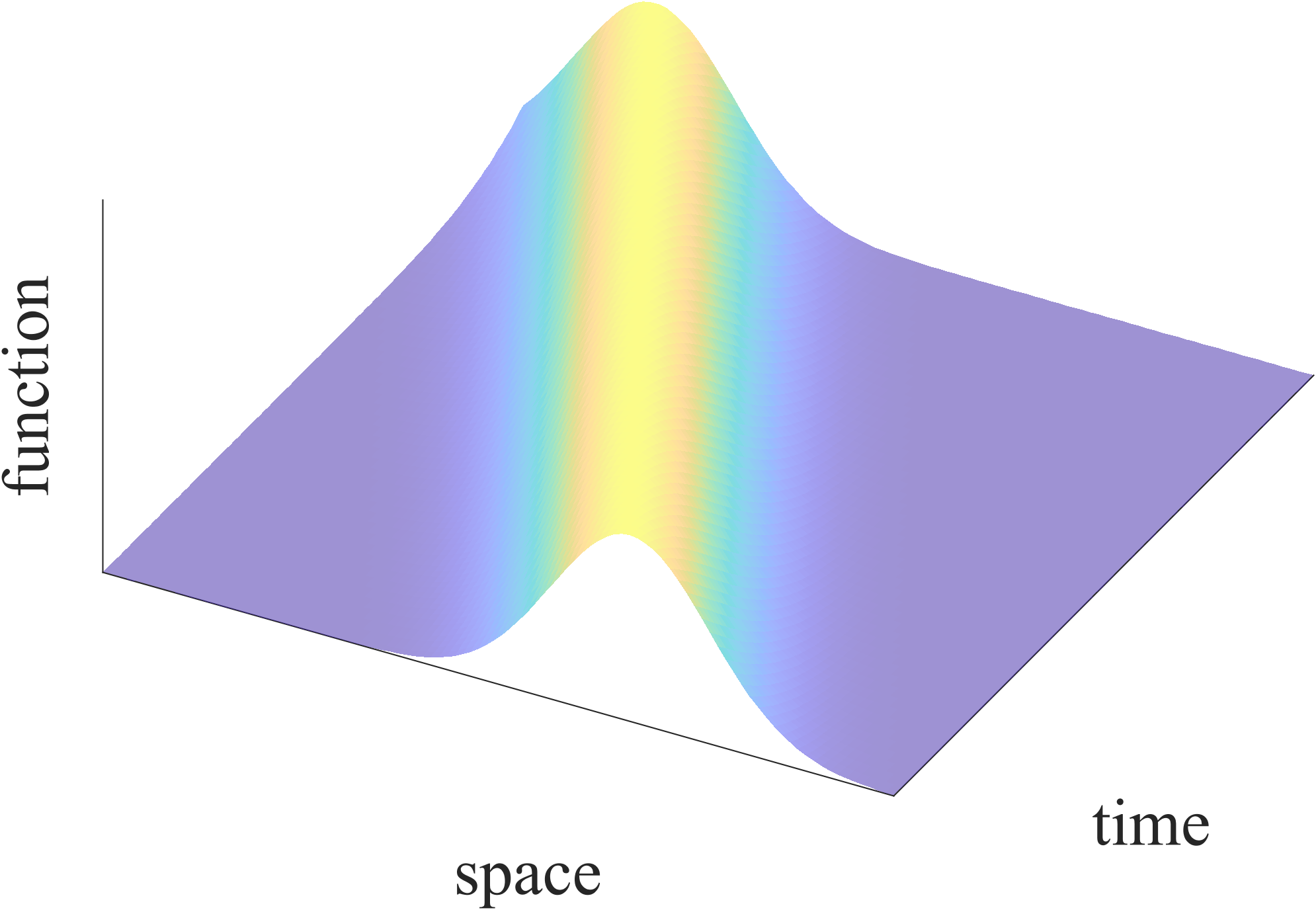}};
	\node[inner sep=0pt, below of=base_plot, xshift=-2cm] (disc_space) {
	    \includegraphics[width=0.43\columnwidth]{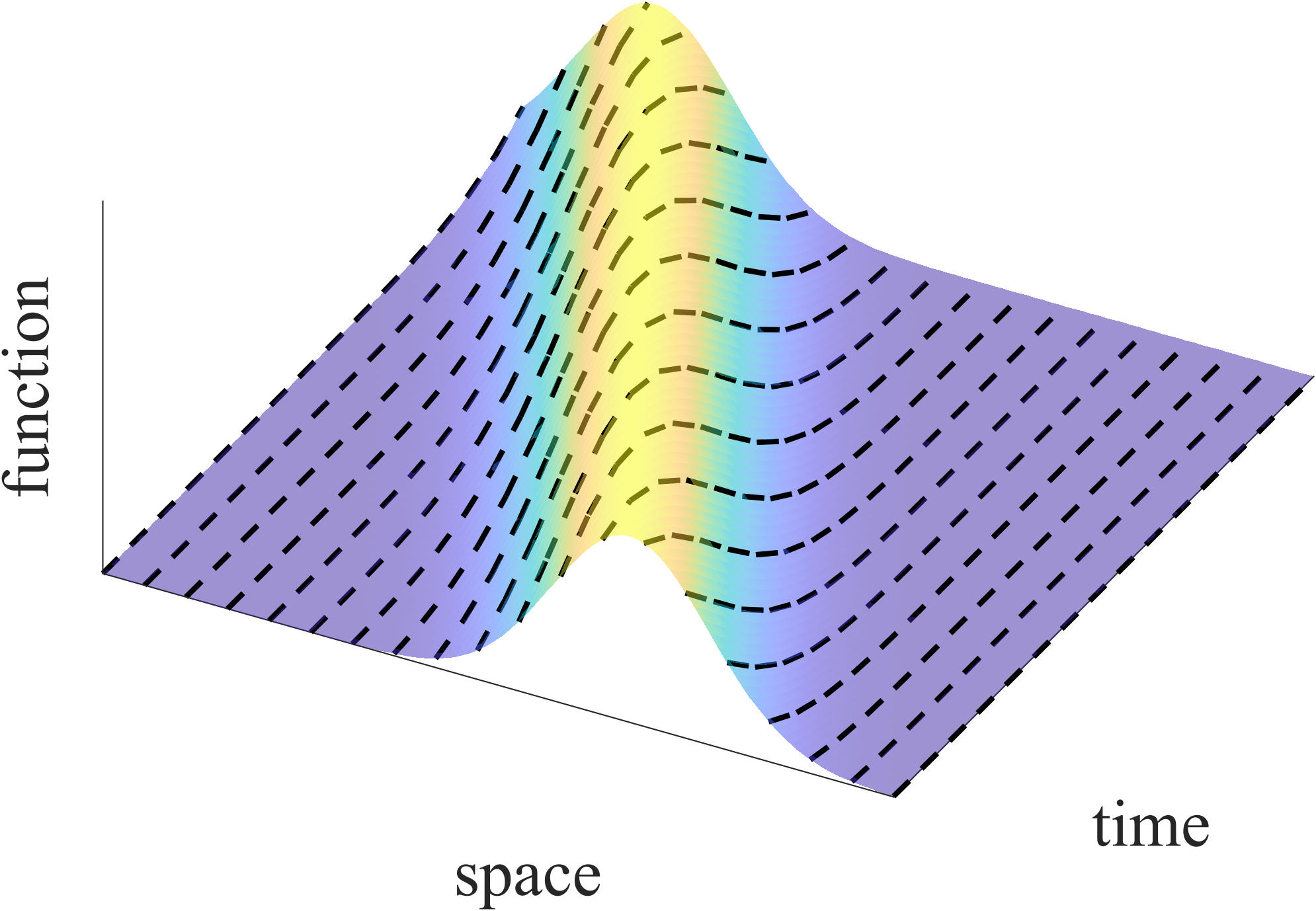}};
	\node[inner sep=0pt, below of=base_plot, xshift=2cm] (disc_time) {
	    \includegraphics[width=0.43\columnwidth]{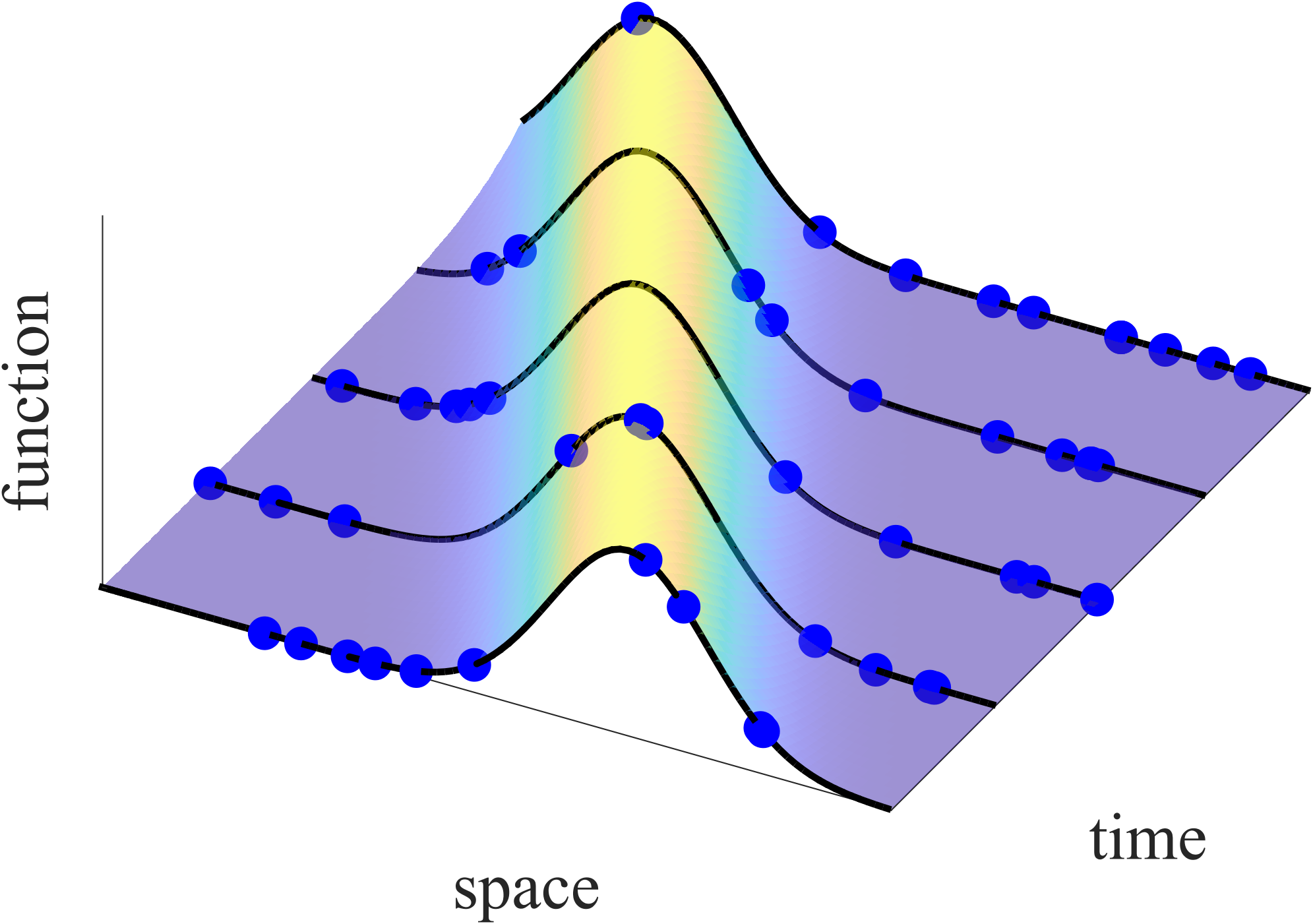}}; 

	\draw [arrow, line width=0.5mm] (base_plot) -- node[midway,anchor=east, align=left] {spatial discretization} (disc_space);
	\draw [arrow, line width=0.5mm] (base_plot) -- node[midway,anchor=west, align=right] {temporal discretization} (disc_time);   
\end{tikzpicture}}
\caption{Spatial discretization methods lay down a mesh on the spatial domain and describe how neighboring mesh cells should interact with each other dictated by the PDE.
This results in a system of ODEs that is simulated with traditional solvers.
Numerical Gaussian processes \cite{raissi2018numerical} on the other hand start with a temporal discretization.
The resulting time discretized PDE is used to build up a structured Gaussian processes that is evaluated on arbitrarily chosen regression points.
To simulate, one recursively formulates the so-called posterior Gaussian process. Figure adapted from \cite{kueper2020mtns}.}
\label{fig:test}
\end{center}
\end{figure}

In this paper, we will take a different route.
We will use so-called numerical Gaussian processes \cite{raissi2018numerical} to solve spatiotemporal models.
Gaussian processes (GPs) are a probabilistic machine learning method that can be used for regression \cite{williams2006gaussian}.
Numerical GPs are structured by the spatiotemporal model in the form of a (linear) time-dependent partial differential equation
\begin{align}
	\dfrac{\partial}{\partial t} f(t,\boldsymbol{x}) &= \mathcal{L}_x f(t,\boldsymbol{x}), \quad \boldsymbol{x} \in \Omega \subset \mathbb{R}^{d_x} \\
	f(t,\boldsymbol{x_b})  &= \mathcal{B}_x f(t,\boldsymbol{x}), \quad \boldsymbol{x_b} \in \partial \Omega,
	\label{eq:linearpde}
\end{align}
where $\mathcal{L}_x: \textit{L}_2 (\mathbb{R}^{d_x},\mathbb{R}) \rightarrow \textit{L}_2 (\mathbb{R}^{d_x},\mathbb{R}),\, f(t,x) \mapsto \mathcal{L}_x f(t,x)$ is a linear (integro-)differential operator determining the dynamics, and $\mathcal{B}_x$ is a linear functional imposing the boundary condition.

Due to the probabilistic nature of GPs, they provide us with uncertainty quantification.
This will be of much benefit for their use in state estimation.

The contributions of this paper are as follows:
\begin{itemize}
	\item we extend numerical GPs with an output channel for online measurements and formulate this as a probabilistic linear state space model which is by definition Gaussian distributed, see Section \ref{sec:nGP2SS};
	\item this allows us to put numerical GPs through the Kalman filter algorithm and therefore derive the numerical Gaussian process Kalman filter (GPKF), see Section \ref{sec:numGPKFderivation};
	\item using numerical case studies of spatiotemporal systems, we showcase the numerical GPKF, see Section \ref{sec:casestudies}.
\end{itemize}

Preliminary results were presented in the conference paper \cite{kueper2020ifacnGPKF}.
This article significantly extends this previous work.
In contrast to \cite{kueper2020ifacnGPKF}, the preliminaries now contain a didactical example to streamline the section.
Furthermore, numerical GPs are explicitly expressed as probabilistic state space models in this paper. 
Moreover, boundary conditions are treated as part of the model and not as measurements.
This affects the KF algorithm as such that boundary conditions enter through the prediction step, rather than the update step of the Kalman filter.
Finally, more sophisticated case studies are presented.

\subsubsection*{Related works}
Infinite-dimensional Kalman filtering with Gaussian processes has previously been done by \cite{sarkka2012infinite,sarkka2013spatiotemporal}.
Therein, certain covariance functions are shown to be convertible via Fourier-transforms to infinite-dimensional state space models.
Using these models, infinite-dimensional Kalman filtering can then be done.
This approach does not need a structured PDE to inform the GPs, contrary to numerical GPs.
Of course, this also means that the GP has to learn the dynamics from the full state vector which might not be accessible in the first place. 

Another fully non-parametric approach is given in \cite{TODESCATO2020109032}.
Assuming space-time separability of the covariance function, a finite-dimensional discrete-time state space model is built up that can be used for Kalman filtering.
Again, the full states need to be accessible for this method to work.    

Finite-dimensional filtering with Gaussian processes has been done by \cite{ko2009gp}.
Different Bayesian filters such as particle filters, extended and unscented Kalman filters are recovered.
Closely related is the work of \cite{deisenroth2009analytic}, wherein an analytic moment-based Gaussian process filter is presented.
A general perspective on finite-dimensional Gaussian filtering is given in \cite{deisenroth2011general}.

\subsubsection*{Mathematical notation}
Scalars are lowercase non-bold symbols $x$, while vectors are bold $\boldsymbol{x}$, and matrices are uppercase and bold $\boldsymbol{X}$.

A random variable $x$ stemming from probability density function $p(\cdot)$ is symbolically written as $x \sim p(x)$.
Gaussian random variables and their distributions are denoted as $x \sim \text{N}(x|m,P)$ with mean $m$ and variance $P$.

A covariance matrix $\boldsymbol{K}\left(\boldsymbol{X}, \boldsymbol{X} \right) \in \mathbb{R}^{M \times M}$ in the Gaussian process regression context is built up by evaluating the underlying covariance function element-wise $\boldsymbol{K}_{ij} = k(\boldsymbol{x}_i,\boldsymbol{x}_j)$ using the data matrix $\boldsymbol{X}= \{ \boldsymbol{x}_1,\dotso,\boldsymbol{x}_M \} \in \mathbb{R}^{d_x \times M}$.
A cross-covariance matrix $\boldsymbol{K} \left( \boldsymbol{X}, \boldsymbol{X}_* \right) \in \mathbb{R}^{M \times S}$ can also be built up with $\boldsymbol{X}_* \in \mathbb{R}^{d_x \times S}$.

\section{Preliminaries}
In this section we will introduce Gaussian process regression, linear operators in combination with GP regression, and numerical GPs for time-dependent partial differential equations.

\subsection*{Gaussian process regression}

A Gaussian process is a stochastic process $f(\boldsymbol{x})$ that is fully defined by its mean function $m (\boldsymbol{x})$ and covariance function $k(\boldsymbol{x},\boldsymbol{x}')$ 
\begin{align}
  m (\boldsymbol{x}) &= \text{E} \left[ f(\boldsymbol{x}) \right] \\
  k(\boldsymbol{x},\boldsymbol{x}') &= \text{E} \left[ \left( f(\boldsymbol{x}) - m(\boldsymbol{x}) \right) \left( f(\boldsymbol{x'}) - m(\boldsymbol{x'}) \right)^T \right].
  \label{eq:gpdefinition}
\end{align} 
What makes GPs usable in practice is their property that any finite dimensional collection of random variables $\boldsymbol{f}(\boldsymbol{X}) = \{ f(\boldsymbol{x}_1),\dotso,f(\boldsymbol{x}_M) \}$ is jointly Gaussian distributed
\begin{align}
\boldsymbol{f}(\boldsymbol{X})
	\sim \text{N} \left(
	\boldsymbol{m}(\boldsymbol{X}),
	\boldsymbol{K}\left(\boldsymbol{X}, \boldsymbol{X} \right)
	\right),
\end{align}
with
\begin{align}
	\boldsymbol{m}(\boldsymbol{X}) &= \begin{pmatrix} m(\boldsymbol{x_1}) & \cdots & m(\boldsymbol{x}_M) \end{pmatrix}^T \\
	\boldsymbol{K}\left(\boldsymbol{X}, \boldsymbol{X} \right) &= 
	\begin{pmatrix}
		k(\boldsymbol{x}_1,\boldsymbol{x}_1) & \cdots & k(\boldsymbol{x}_1,\boldsymbol{x}_M)  \\ 
		\vdots & \ddots & \vdots \\
		k(\boldsymbol{x}_M,\boldsymbol{x}_1) & \cdots & k(\boldsymbol{x}_M,\boldsymbol{x}_M) 
	\end{pmatrix}.
\end{align}

In regression, we wish to infer the process $f(\cdot)$ based on its inputs $\boldsymbol{x}$ and its (possibly noisy) output $y$.
The output is not restricted to algebraic expressions such as $y = f(\boldsymbol{x}) + noise$, but can also be generalized to linear transforms of the process $y = \mathcal{L}_x f(\boldsymbol{x}) + noise$.

\begin{figure}[tb!]
\centering
\begin{subfigure}[t]{0.48\columnwidth}
		\includegraphics[width=1\columnwidth]{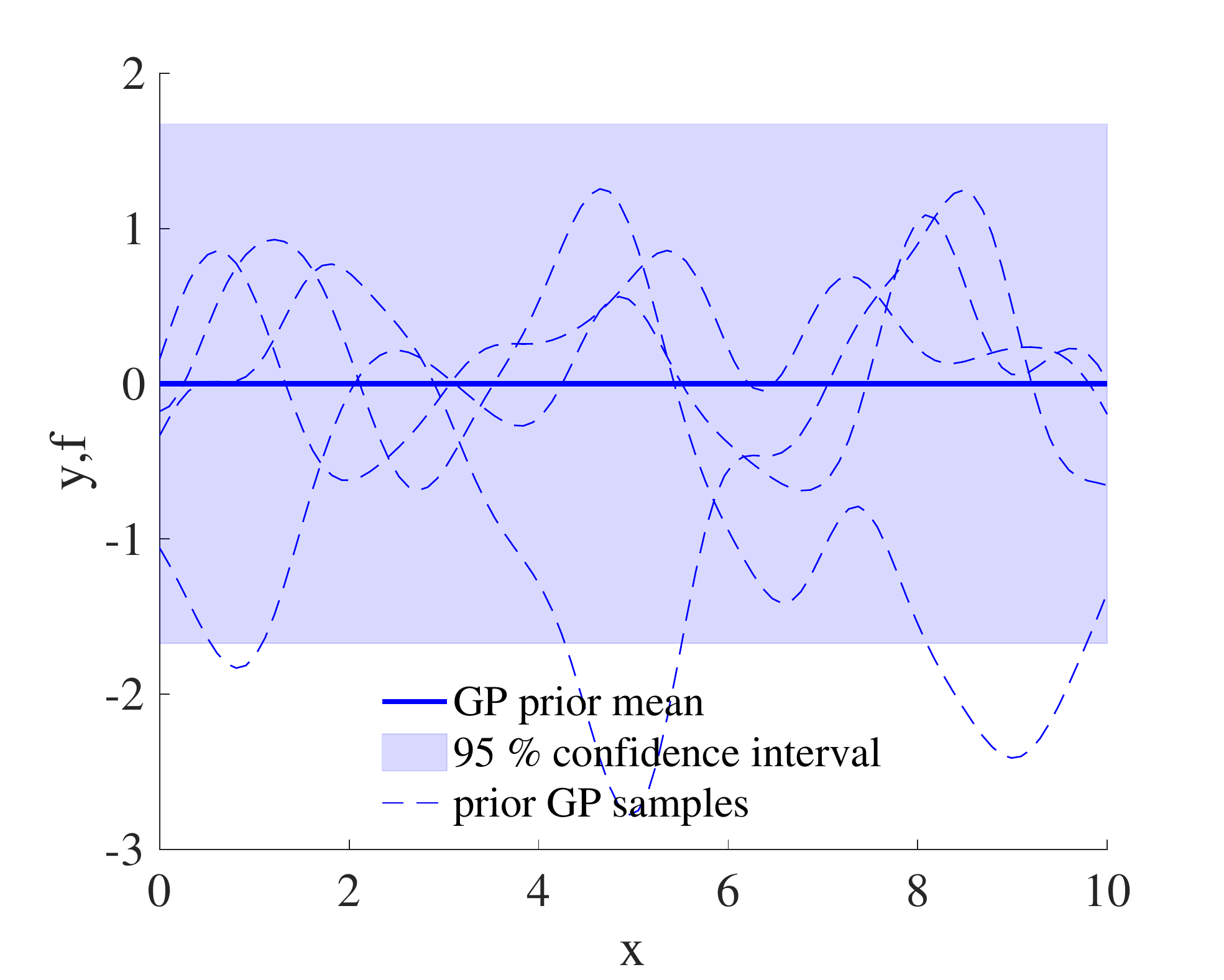}
		\label{subfig:GP_differential_inference_prior}
		\caption{Gaussian process prior consisting of a zero mean and the squared exponential covariance function.}
	\end{subfigure} \hspace*{0.15cm}%
	\begin{subfigure}[t]{0.48\columnwidth}
		\includegraphics[width=1\columnwidth]{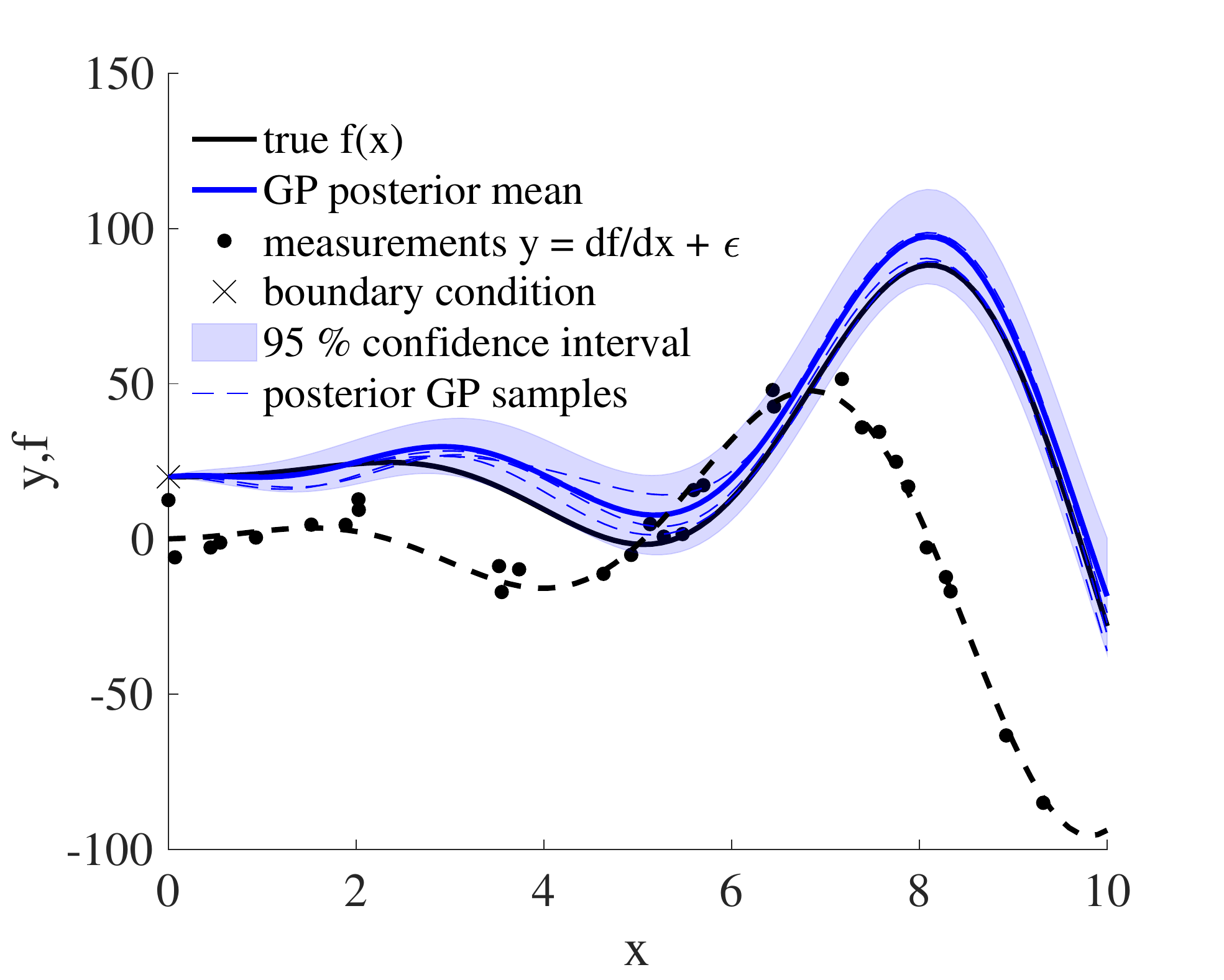}
		\label{subfig:GP_differential_inference_post}
		\caption{Corresponding posterior Gaussian process.}
	\end{subfigure}
	\caption{Inferring $f(x)$ from noisy differential observations $y = \tfrac{\text{d} f}{\text{d} x} + \epsilon$ and boundary condition $f^b=f(x_b)$ using Gaussian process regression.
	The true function is $f_{true} = x^2 \sin (x) + 0.2 x^{3/2} + 20$.} 
	\label{fig:gp_differential_inference}
\end{figure}

We will illustrate GP regression using the following differential example 
\begin{align}
y &= \dfrac{\text{d} f}{\text{d} x} + \epsilon, \quad \text{with} \: \epsilon \sim \text{N} \left(0, \sigma_{\epsilon}^2 \right), \\
f^b &= \mathcal{B}_x f,
\label{eq:GPdifferentialexample}
\end{align}
with a Dirichlet boundary condition rendering the boundary operator to the identity operator $\mathcal{B}_x = \mathcal{I}_x$, and therefore  $f^b=f(x_b=0)$. 
To start, we place a GP prior on $f(x)$ and write
\begin{align}
	f(x) \sim \text{GP} \left(0, k^{ff}(x,x') \right),
\end{align}
where we chose the prior consisting of a zero-mean function and a squared exponential covariance function $k(x,x') = \sigma^2 \exp \left( - \tfrac{(x-x')^2}{2 l^2} \right)$.
Here, $\sigma^2$ and $l$ are hyper-parameters that permit to adjust the shape of the covariance function, and therefore the behavior of $f(x)$, to the data.

Next, we make use of the fact that a linear transform of a GP is another GP that is structured by the linear transform, see \cite{sarkka2011linear} and the references therein.
We therefore have
\begin{align}
y &\sim \text{GP} \left( 0, \underbrace{\mathcal{L}_x \mathcal{L}_{x'} k^{ff}(x,x')}_{=k^{yy}} \right), \\
f^b &\sim \text{GP} \left( 0, \underbrace{\mathcal{B}_x \mathcal{B}_{x'} k^{ff}(x,x')}_{=k^{bb}} \right).
\end{align}
For our example, $k^{ff}$ and $k^{bb}$ are identical due to the Dirichlet boundary condition. 
The covariance function $k^{yy}$ can easily be derived under use of the formal definition of a covariance function \eqref{eq:gpdefinition}.
Note the important distinction between $\mathcal{L}_x$ and $\mathcal{L}_{x'}$, meaning that the operator should be applied with respect to either the first or second argument, that is $x$ or $x'$.

We continue by formulating the evaluated joint model $p(\boldsymbol{f},f^b,\boldsymbol{y})$ at test locations $\boldsymbol{x}_* = \{x_{*1},\hdots,x_{*S} \}$, the boundary $x_b$, and observation locations $\boldsymbol{x} = \{ x_1, \hdots, x_{M} \}$
\begin{align}
	\begin{pmatrix}
		\boldsymbol{f} (\boldsymbol{x}_*)  \\
		f^b (x_b) \\
		\boldsymbol{y} (\boldsymbol{x})
	\end{pmatrix}
	\sim \text{N}
	\left( 
	\boldsymbol{0},
	\boldsymbol{K}
	\right),
	\label{eq:gpjointmodel_example}
\end{align}
with covariance matrix 
\begin{align}
\boldsymbol{K} = 
\begin{bmatrix}
	\boldsymbol{K}^{ff} \left( \boldsymbol{x}_*, \boldsymbol{x}_* \right) & \boldsymbol{k}^{fb} \left( \boldsymbol{x}_*, x_b \right) & \boldsymbol{K}^{fy} \left( \boldsymbol{x}_*, \boldsymbol{x} \right) \\
	\boldsymbol{k}^{bf} \left( x_b, \boldsymbol{x}_* \right) & k^{bb} (x_b, x_b) & \boldsymbol{k}^{by} \left( x_b, \boldsymbol{x} \right)\\
	\boldsymbol{K}^{yf} \left( \boldsymbol{x}, \boldsymbol{x}_* \right) & \boldsymbol{k}^{yb} \left( \boldsymbol{x}, x_b \right) & \boldsymbol{K}^{yy} \left( \boldsymbol{x}, \boldsymbol{x} \right)
	\end{bmatrix}.
\end{align}

Before doing inference, we learn the hyper-parameters of the covariance function by minimizing the negative log marginal likelihood (NLML) $- \log p(f^b,\boldsymbol{y})$.
We will touch on this more thoroughly later on.

Inference is now done by calculating the conditional distribution $p(\boldsymbol{f} | f^b, \boldsymbol{y})$.
For Gaussian distributions, this results in another Gaussian with conditional mean $\text{E} [\boldsymbol{f} | f^b, \boldsymbol{y}]$ and covariance $\text{V} [\boldsymbol{f} | f^b, \boldsymbol{y}]$ as
\begin{align}
\text{E} [\boldsymbol{f} | f^b, \boldsymbol{y}] =& \begin{bmatrix} \boldsymbol{k}^{fb} & \boldsymbol{K}^{fy} \end{bmatrix}
\begin{bmatrix} k^{bb} & \boldsymbol{k}^{by} \\
 \boldsymbol{k}^{yb} & \boldsymbol{K}^{yy} + \sigma_n^2 \boldsymbol{I} \end{bmatrix} ^{-1} \begin{pmatrix} f^b \\ \boldsymbol{y} \end{pmatrix} \\
\text{V} [\boldsymbol{f} | f^b, \boldsymbol{y}] =&\: \boldsymbol{K}^{ff} \nonumber \\
&-\: \begin{bmatrix} \boldsymbol{k}^{fb} & \boldsymbol{K}^{fy} \end{bmatrix} \begin{bmatrix} k^{bb} & \boldsymbol{k}^{by} \\
 \boldsymbol{k}^{yb} & \boldsymbol{K}^{yy} + \sigma_n^2 \boldsymbol{I} \end{bmatrix} ^{-1} \begin{bmatrix} \boldsymbol{k}^{bf} \\ \boldsymbol{K}^{yf} \end{bmatrix}.
 \label{eq:condmeanandcov}
\end{align} 
Here, another hyper-parameter $\sigma_n^2$ has been introduced to model measurement noise $\epsilon$ as additive zero mean Gaussian white noise.
This hyper-parameter is also learned via the NLML.
As can be seen in Fig. \ref{fig:gp_differential_inference}, the structured GP \eqref{eq:gpjointmodel_example} is able to infer $f$ from derivative observations.  

To summarize, we give a cooking recipe for probabilistic inference with GPs:
\begin{enumerate}
	\item Carefully place a GP prior to prevent inversion of linear operators (prior on $f$ in our example to prevent inversion of $\tfrac{\text{d}}{\text{d} x}$).
	\item Build up output GPs that are linear transforms of the GP prior ($y,f^b$ in our example). 
	\item Construct the joint model $p(f,f^b,y)$ and learn the hyper-parameters via the marginal distribution $p(y,f^b)$.
	\item Do inference by conditioning $p(f|f^b,y)$.
\end{enumerate}

\subsection*{Numerical Gaussian processes for time-dependent partial differential equations}

So far we looked at static examples.
We are however interested in spatiotemporal systems described through time-dependent partial differential equations (PDEs).
We assume that the PDE model is known and linear.
Given an initial condition and boundary conditions, we wish to solve the PDE via GPs.
We can do this with the recently introduced numerical GPs \cite{raissi2018numerical}.
We will present the essential idea here, referring the reader to the original paper for further details.

Numerical GPs are built on the fact that linear transforms of a GP result in another GP.
In fact, numerical GPs use a very similar procedure as shown earlier.

The first step is to discretize the time-dependent PDE \eqref{eq:linearpde} in time using one of the many existing methods (explicit Euler for simplicity here)
\begin{align}
	f_t(\boldsymbol{x}) &= f_{t-1}(\boldsymbol{x}) + \Delta t \mathcal{L}_x f_{t-1}(\boldsymbol{x}) \\
	&= \mathcal{A}_x f_{t-1}(\boldsymbol{x}).
\end{align}
Next, a GP prior is placed on $f_{t-1} \sim \text{GP} \left(0, k^{ff}_{t-1,t-1} (\boldsymbol{x},\boldsymbol{x}') \right)$ and therefore $f_t$ is also a GP defined as
\begin{align}
	f_t(\boldsymbol{x}) \sim \text{GP} \left(0, \underbrace{\mathcal{A}_x \mathcal{A}_{x'} k^{ff}_{t-1,t-1}}_{= k^{ff}_{t,t}} \right),
\end{align}
and so is the boundary condition $f^b_t = \mathcal{B}_x f_t$
\begin{align}
	f^b_t(\boldsymbol{x}) \sim \text{GP} \left(0, \underbrace{\mathcal{B}_x \mathcal{B}_{x'} \mathcal{A}_x \mathcal{A}_{x'} k^{ff}_{t-1,t-1}}_{= k^{f^b f^b}_{t,t}} \right).
\end{align}
Then, the evaluated joint model $p(\boldsymbol{f}_t, \boldsymbol{f}^b_t, \boldsymbol{f}_{t-1})$ on arbitrary test points $\boldsymbol{X}_t$, the boundaries $\boldsymbol{X}_b$, and previous points $\boldsymbol{X}_{t-1}$ is formulated as 
\begin{align}
	\begin{pmatrix}
		\boldsymbol{f}_t \\
		\boldsymbol{f}^b_t \\
		\boldsymbol{f}_{t-1}
	\end{pmatrix}
	\sim \text{N}
	\left( 
	\boldsymbol{0},
	\begin{bmatrix}
	\boldsymbol{K}^{ff}_{t,t} & \boldsymbol{K}^{ff^b}_{t,t} & \boldsymbol{K}^{ff}_{t,t-1} \\
	\boldsymbol{K}^{f^b f}_{t,t} & \boldsymbol{K}^{f^b f^b}_{t,t} & \boldsymbol{K}^{f^b f}_{t,t-1}\\
	\boldsymbol{K}^{ff}_{t-1,t} & \boldsymbol{K}^{ff^b}_{t-1,t} & \boldsymbol{K}^{ff}_{t-1,t-1} 
	\end{bmatrix}
	\right).
	\label{eq:numgpjointmodel_example}
\end{align}
Here, the inputs to the individual covariance matrices are omitted.
They can be inferred from the sub- and superscripts, e.g. $\boldsymbol{K}^{ff^b}_{t-1,t}(\boldsymbol{X}_{t-1},\boldsymbol{X}_b)$.

To simulate the spatiotemporal model, we now recursively compute the conditional distribution $p(\boldsymbol{f}_t | \boldsymbol{f}^b_t, \boldsymbol{f}_{t-1})$.
Here the solution from the previous simulation step $\boldsymbol{f}_{t-1}$ acts as an artificial measurement.
Hyper-parameters are optimized in each simulation step by minimizing $-\log p(\boldsymbol{f}^b_t, \boldsymbol{f}_{t-1})$.
\section{Numerical Gaussian processes as probabilistic state space models}
\label{sec:nGP2SS}

In this section we will derive a probabilistic state space model from numerical Gaussian processes.
This probabilistic state space model is linear and its random states are, by definition of GPs, Gaussian distributed.
These model properties allow us in Section \ref{sec:numGPKFderivation} to go through the Kalman filter algorithm, thereby deriving the numerical Gaussian process Kalman filter.

We start by introducing an output channel for noisy online measurements that are linear transforms of $f(t,\boldsymbol{x})$.
Mathematically this reads as
\begin{align}
	f^y_t( \boldsymbol{y}) &= \left( \mathcal{H}_x f_t \right) ( \boldsymbol{x})  + r_t( \boldsymbol{y}).
\end{align}
The measurement operator is $\mathcal{H}_x: \textit{L}_2 (\mathbb{R}^{d_x},\mathbb{R}) \rightarrow \textit{L}_2 (\mathbb{R}^{d_y},\mathbb{R}),\, f_t \mapsto f^y_t$.
Obtained measurements are point evaluations of $f^y_t$ at locations $\boldsymbol{y}_i$.

In this work, we consider measurement noise $r_t$ as zero-mean spatiotemporal white noise with constant variance over $\boldsymbol{y}$.
To model this, we place a GP prior on $r_t$ that is independent of all the other GP priors.
We write
\begin{align}
r_t \sim \text{GP} \left( 0, k^{rr}_{t,t} (\boldsymbol{y},\boldsymbol{y}') \right),
\end{align}
where the covariance function is $k^{rr}_{t,t} = \sigma_r^2 \delta(\boldsymbol{y}-\boldsymbol{y}')$ and $E [r_t r_{t+\tau}]=0$.
Other additive noise models are conceivable.
Particularly the spatial white noise assumption could be relaxed to something more suitable under use of an appropriate covariance function \cite{williams2006gaussian}.

Since $f_t(x)$ is a GP, so is the measurement output
\begin{align}
f^y_t \sim \text{GP} \left( 0, \underbrace{\left( \mathcal{H}_x \mathcal{H}_{x'} k^{ff}_{t,t} \right) (\boldsymbol{x}, \boldsymbol{x}') + k^{rr}_{t,t} (\boldsymbol{y}, \boldsymbol{y}')}_{=k^{f^y f^y}_{t,t}} \right).
\end{align}

Regarding the PDE itself, we add spatiotemporal process noise to the PDE to obtain
\begin{align}
  \dfrac{\partial}{\partial t} f(t,\boldsymbol{x}) &= \mathcal{L}_x f(t,\boldsymbol{x}) + q(t,\boldsymbol{x}).
  \label{eq:SPDE_spatiotemporalnoise}
\end{align}
Discretization in time yields
\begin{align}
	f_t(\boldsymbol{x}) &= \mathcal{A}_x f_{t-1}(\boldsymbol{x}) + \Delta t q_{t-1}(\boldsymbol{x}),
\end{align}
with process noise modeled as an independent zero-mean GP prior with a white noise kernel $k^{qq}_{t-1,t-1} (x,x') = \sigma_q^2 \delta(x-x')$.
Of course, the above discretization is mathematically not rigorous because \eqref{eq:SPDE_spatiotemporalnoise} is not continuous due to the spatiotemporal white noise.
A more rigorous treatment would require It{\^o}--Calculus \cite{sarkka2019applied}.
We could formally circumvent this by first discretizing and then adding process noise $q_{t-1}(\boldsymbol{x})$ (without $\Delta t$), as done in the finite-dimensional state space models.
However, we found that exclusion of $\Delta t$ in the process noise term drastically reduced estimation performance.  
Moreover, above we recovered a time-discrete form that is similar to the Euler-Maruyama method.
This brief discussion lays bare the gap between proper theoretical treatment and actual implementation, as lamented as early as in \cite{curtain1975survey}.

The updated GP for $f_t(x)$ is
\begin{align}
	f_t(x) \sim \text{GP} \left(0, \underbrace{\mathcal{A}_x \mathcal{A}_{x'} k^{ff}_{t-1,t-1} +  \Delta t^2 k^{qq}_{t-1,t-1}}_{=k^{ff}_{t,t}} \right).
\end{align}

The complete multi output GP now reads as
\begin{equation}
\begin{pmatrix}
  f_{t}\\
  f^y_t \\
  f_t^b \\
  f_{t-1}
\end{pmatrix}
\sim \mathrm{GP} \left(
\boldsymbol{0},
\underbrace{
\begin{bmatrix}
k_{t,t}^{ff} & k_{t,t}^{f^y f} & k^{f f^b}_{t,t} & k_{t,t-1}^{ff} \\
k_{t,t}^{f^y f} & k_{t,t}^{f^y f^y} & k^{f^y f^b}_{t,t} & k_{t,t-1}^{f^y f} \\
k^{f^b f}_{t,t} & k^{f^b f^y}_{t,t} & k^{f^b f^b}_{t,t} & k^{f^b f}_{t,t-1} \\ 
k_{t-1,t}^{ff} & k_{t-1,t}^{ff^y} & k^{f f^b}_{t-1,t} & k_{t-1,t-1}^{ff} 
\end{bmatrix}}_{=\boldsymbol{K}_{GP}}
\right).
\label{eq:multioutputGPfull}
\end{equation}
All covariance functions inside $\boldsymbol{K}_{GP}$ are known as they can be built up from the prior covariance function $k^{ff}_{t-1,t-1}$ (for the explicit Euler)  and the respective linear operators $\mathcal{A}_x,\, \mathcal{H}_x$ or functionals $\mathcal{B}_x$, see Appendix \ref{app:nGPSSprecursor}.

We can recover a probabilistic state space model from \eqref{eq:multioutputGPfull}.
For this purpose we assume that \eqref{eq:multioutputGPfull} has been evaluated at measurement locations $\boldsymbol{Y}$, boundaries $\boldsymbol{X}_b$ and arbitrarily chosen test locations $\boldsymbol{X}$.
The covariance functions therefore become covariance matrices, e.g. $k^{f^y f^b}_{t,t}(\boldsymbol{y},\boldsymbol{x}_b)$ becomes $\boldsymbol{K}^{f^y f^b}_{t,t}(\boldsymbol{Y},\boldsymbol{X}_b)$. 

The process equation is
\begin{align}
p \left( \boldsymbol{f}_t| \boldsymbol{f}^b_t, \boldsymbol{f}_{t-1} \right)
&= \mathrm{N} \left(
  \boldsymbol{f}_t \big\vert
  \boldsymbol{A}_t 
  \begin{pmatrix} \boldsymbol{f}^b_t \\ \boldsymbol{f}_{t-1} \end{pmatrix},\:
 \boldsymbol{P}^{GPf}_t
\right),
\label{eq:nGPprocess}
\end{align}
with the covariance matrix
\begin{align}
\boldsymbol{A}_t =& \begin{bmatrix}\boldsymbol{K}^{ff^b}_{t,t} & \boldsymbol{K}^{ff}_{t,t-1} \end{bmatrix} 
  \begin{bmatrix} \boldsymbol{K}^{f^b f^b}_{t,t} & \boldsymbol{K}^{f^b f}_{t,t-1} \\
  \boldsymbol{K}^{ff^b}_{t-1,t} & \boldsymbol{K}^{ff}_{t-1,t-1} \end{bmatrix}^{-1} 
\end{align}
taking the role of the dynamic matrix in a state space model.
Here, $\boldsymbol{A}_t$ is time-varying, indicated by the subscript, due to changing hyper-parameter values and possibly changing regression points.
The process noise covariance matrix of the state space model, usually denoted $\boldsymbol{Q}$, is given by
 \begin{align}
\boldsymbol{P}^{GPf}_t = \boldsymbol{K}^{ff}_{t,t} - \boldsymbol{A}_t \begin{bmatrix}\boldsymbol{K}^{ff^b}_{t,t} & \boldsymbol{K}^{ff}_{t,t-1} \end{bmatrix}^T.
 \end{align}

The measurement equation reads as
\begin{align}
p \left( \boldsymbol{f}^y_t | \boldsymbol{f}_t \right)
&= \mathrm{N} \left(
  \boldsymbol{f}^y_t \bigg\vert 
  \boldsymbol{C}_t \boldsymbol{f}_{t}, \: \boldsymbol{P}^{GP,f^y}_t 
  \right),
\label{eq:nGPmeasurement}
\end{align}
with
\begin{align}
\boldsymbol{C}_t = \boldsymbol{K}^{f^y f}_{t,t} \left( \boldsymbol{K}^{ff}_{t,t} \right)^{-1}
\end{align}
taking the role of the measurement matrix in a state space model.
The measurement noise covariance matrix of the state space model, usually denoted $\boldsymbol{R}$, is given by
\begin{align}
  \boldsymbol{P}^{GP,f^y}_t = \boldsymbol{K}^{f^y f^y}_{t,t} - \boldsymbol{C}_t \boldsymbol{K}^{ff^y}_{t,t}.
\end{align}

The noise covariance matrices $\boldsymbol{Q}$ and $\boldsymbol{R}$ in a classical state space model are usually static matrices.
The GP model however naturally allows for an adaptive noise description, due to online adaptation of the hyper-parameters, thereby drastically reducing the usually required tedious fine-tuning of the Kalman filter.
This will be showcased with case studies in Section~\ref{sec:casestudies} and discussed in Section~\ref{sec:discussion}.

For the recursive Kalman filter algorithm we will also need the following 
\begin{align}
p(\boldsymbol{f}^b_t | \boldsymbol{f}_{t-1}) = \mathrm{N} \left( \boldsymbol{f}^b_t | \boldsymbol{A}^{f^b}_t \boldsymbol{f}_{t-1}, \boldsymbol{P}^{GP,f^b}_t \right),
\end{align}
with
\begin{align}
\boldsymbol{A}^{f^b}_t \:{=}&\: \boldsymbol{K}^{f^b f}_{t,t-1} \left( \boldsymbol{K}^{nn}_{t-1,t-1} \right)^{-1}, \\
\boldsymbol{P}^{GP,f^b}_t \:{=}&\: \boldsymbol{K}^{f^b f^b}_{t,t} - \boldsymbol{A}^{f^b}_t \boldsymbol{K}^{f f^b}_{t-1,t}.
\label{eq:forwardboundarymodel}
\end{align}

Equipped with a probabilistic state space model, we are now ready to write down the recursive Kalman filter equations.
\section{Numerical Gaussian process Kalman filter derivation}
\label{sec:numGPKFderivation}
In this section we will embed the numerical GP state space model into the Kalman filter algorithm.
Afterwards, we will briefly discuss the Kalman filter marginal likelihood for the use of hyper-parameter estimation.

There are two ways to derive the Kalman filter.
The first derivation is rooted in recursive least squares regression, see e.g. \cite{simon2006optimal}.
The second derivation takes a Bayesian filtering perspective, see e.g. \cite{sarkka2013bayesian}.
In this article, we take the latter perspective to derive the numerical Gaussian process Kalman filter since the Bayesian perspective neatly fits into the Gaussian process framework.

Recall that in Kalman filtering, we want to calculate the probability distribution of a dynamic state given all the measurement history up to the present.
For spatiotemporal systems, we additionally condition on the boundary data history.
Boundary conditions are treated as part of the model and therefore enter through the prediction step of the Kalman filter.
According to Bayes' rule, the posterior distribution of the state estimate is given by
\begin{equation}
p(\boldsymbol{f}_t|\boldsymbol{f}^y_{1:t}, \boldsymbol{f}^b_{1:t}) = \dfrac{p(\boldsymbol{f}^y_t | \boldsymbol{f}_t) p(\boldsymbol{f}_t|\boldsymbol{f}^y_{1:t-1}, \boldsymbol{f}^b_{1:t})}{p(\boldsymbol{f}^y_t | \boldsymbol{f}^y_{1:t-1}, \boldsymbol{f}^b_{1:t})}.
\label{eq:numericalGPKFBayes}
\end{equation}

We make the usual assumptions for the Kalman filter \cite{sarkka2013bayesian} and an additional one for the boundary data:
\begin{enumerate}[
labelindent=*,
style=multiline,
leftmargin=*,
label=Assumption \arabic*)
]
  \item States are assumed to be Markovian, i.e. the current state $\boldsymbol{f}_t$ is conditionally independent of anything that happened before $t-1$ given the previous state $\boldsymbol{f}_{t-1}$.
  \item Given the current state $\boldsymbol{f}_t$, the current measurement $\boldsymbol{f}^y_t$ is conditionally independent of the measurement $\boldsymbol{f}^y_{1:t-1}$ and state histories $\boldsymbol{f}_{1:t-1}$, as well as the boundary history up to the present $\boldsymbol{f}^b_{1:t}$.
  \item Given the previous state $\boldsymbol{f}_{t-1}$, the current boundary condition $\boldsymbol{f}^b_t$ is conditionally independent of the measurement $\boldsymbol{f}^y_{1:t-1}$ and boundary history data $\boldsymbol{f}^b_{1:t-1}$.
\end{enumerate}
In this article, only one-step discretization schemes such as the Euler method are considered.
Using multi-step methods would require to extend the Markovian property to an accordingly higher order.

\begin{figure}[tb]
  \begin{center}
  \resizebox{1\columnwidth}{!}{%
    \begin{tikzpicture}[node distance=2cm]
      \node (priormodel) [block] {$p(f_t|f^b_t, f_{t-1})$};
      \node (times1) [operation, right of=priormodel] {$\times$};
      \node (previouspost) [block, right of=times1, xshift=1cm] {$p(f_{t-1}|f^y_{1:t-1}, f^b_{1:t-1})$}; 
      \node (priorboundary) [block, above of = times1] {$p(f^b_t | f_{t-1})$};     
      \node (priorjoint) [block, below of=times1] {$p(f_t, f^b_t, f_{t-1}|f^y_{1:t-1}, f^b_{1:t-1})$};
      \node (prior) [blockspecial, below of=priorjoint, yshift=-0.5cm] {\textbf{prior} $p(f_t | f^y_{1:t-1}, f^b_{1:t})$};
      \node (times2) [operation, right of=prior, xshift=7cm] {$\times$};
      \node (likelihood) [blockspecial, below of=times2, yshift=0.5cm] {\textbf{likelihood} $p(f^y_t | f_t)$};
      \node (posteriorjoint) [block, above of=times2] {$p(f_t, f^y_t | f^y_{1:t-1}, f^b_{1:t})$};
      \node (posterior) [blockspecial, above of =posteriorjoint, yshift=0.5cm] {\textbf{posterior} $p(f_t|f^y_{1:t}, f^b_{1:t})$};
      \node (initial) [block, above of=posterior] {$p(f_0)$};

      \draw [arrow] (priormodel) -- (times1);
      \draw [arrow] (previouspost) -- (times1);
      \draw [arrow] (priorboundary) -- (times1);
      \draw [arrow] (times1) -- (priorjoint);
      \draw [arrow] (priorjoint) -- node[anchor=east, align=left] {first marginalize $f_{t-1}$, \\ then condition on $f^b_t$} (prior);
      \draw [arrow] (prior) -- (times2);
      \draw [arrow] (likelihood) -- (times2);
      \draw [arrow] (times2) -- (posteriorjoint);    
      \draw [arrow] (posteriorjoint) -- node[anchor=east] {condition on $f^y_t$} (posterior);  
      \draw [arrow] (posterior) -- node[anchor=south, text width=1.5cm] {repeat recursively $t \rightarrow t-1$} (previouspost);   
      \draw [arrow] (initial) -- node[anchor=west] {start recursion} (posterior);   

      \begin{pgfonlayer}{background}
        \filldraw [fill=black!5,draw=black, rounded corners=2ex]
                (-1.4,2.6) rectangle (6.7,-5.1);
        \filldraw [fill=black!5,draw=black, rounded corners=2ex]
                (6.9,2.6) rectangle (13.3,-6.7);
        \node [right of = priorboundary, xshift = 0.6cm] {PREDICTION STEP};
        \node [left of = initial, xshift = -0.5cm] {UPDATE STEP};
      \end{pgfonlayer} 
    \end{tikzpicture}
    }
    \caption{The recursive Kalman filter algorithm for calculating the posterior state distribution of a spatiotemporal system.}
    \label{fig:recursivefilter}
  \end{center}
\end{figure}
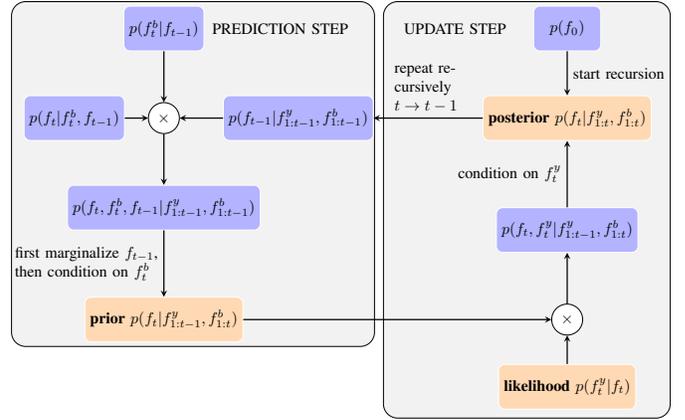

The individual terms of \eqref{eq:numericalGPKFBayes} can be calculated in closed form for Gaussian distributions as
\begin{itemize}
  \item prior \( p(\boldsymbol{f}_t| \boldsymbol{f}_{1:t-1}^y, \boldsymbol{f}^b_{1:t}) = \mathrm{N}(\boldsymbol{f}_t|\boldsymbol{m}^-_t, \boldsymbol{P}^-_t) \)
  \item posterior \( p(\boldsymbol{f}_t|\boldsymbol{f}^y_{1:t}, \boldsymbol{f}^b_{1:t}) = \mathrm{N}(\boldsymbol{f}_t|\boldsymbol{m}_t, \boldsymbol{P}_t) \)
  \item marginal likelihood \\ \( p(\boldsymbol{f}^y_t | \boldsymbol{f}^y_{1:t-1}, \boldsymbol{f}^b_{1:t}) = \mathrm{N}(\boldsymbol{f}^y_t| \boldsymbol{C}_t \boldsymbol{m}^-_t, \boldsymbol{S}_t) \)
\end{itemize}
The road map ahead to derive these expressions is shown in Fig.~\ref{fig:recursivefilter}.
To calculate the prior distribution $p(\boldsymbol{f}_t | \boldsymbol{f}^y_{1:t-1}, \boldsymbol{f}^b_{1:t})$, we will first formulate the joint distribution between states $\boldsymbol{f}_t$, $\boldsymbol{f}_{t-1}$, and current boundary condition $\boldsymbol{f}^b_t$, conditioned on measurement and boundary histories.
Under use of Assumptions 1 and 3 \footnote{regarding the boundary, one could also think of assuming that the current boundary $f^b_t$ and the previous state $f_{t-1}$ are independent but this would be in conflict with the numerical GP state space model.}, the prior joint is calculated as
\begin{align}
  {}&\:p(\boldsymbol{f}_t, \boldsymbol{f}^b_{t}, \boldsymbol{f}_{t-1}|\boldsymbol{f}^y_{1:t-1}, \boldsymbol{f}^b_{1:t-1}) \\
  {=}&\: \underbrace{ p(\boldsymbol{f}_t| \boldsymbol{f}^b_{t}, \boldsymbol{f}_{t-1}) }_{\text{model prediction}} p(\boldsymbol{f}^b_{t} | \boldsymbol{f}_{t-1}) \underbrace{p(\boldsymbol{f}_{t-1}|\boldsymbol{f}^y_{1:t-1}, \boldsymbol{f}^b_{1:t-1})}_{\text{previous posterior}} \\
  {=}&\: \mathrm{N} \left( \boldsymbol{f}_t | \boldsymbol{A}_t \begin{pmatrix} \boldsymbol{f}^b_t \\ \boldsymbol{f}_{t-1} \end{pmatrix}, \boldsymbol{P}^{GP,f}_t \right) \\
  &{\times}\: \mathrm{N} \left( \boldsymbol{f}^b_t | \boldsymbol{A}^{f^b}_t \boldsymbol{f}_{t-1}, \boldsymbol{P}^{GP,f^b}_t \right) \nonumber \\
  &{\times}\: \mathrm{N}\left( \boldsymbol{f}_{t-1}| \boldsymbol{m}_{t-1}, \boldsymbol{P}_{t-1} \right) \nonumber \\
  {=}&\: \mathrm{N} \left( \begin{pmatrix}
  \boldsymbol{f}_t \\
  \boldsymbol{f}^b_{t} \\
  \boldsymbol{f}_{t-1}
  \end{pmatrix} | \boldsymbol{m}', \boldsymbol{P}' \right). 
  \label{eq:priorjoint}
\end{align}
The covariance matrices $\boldsymbol{A}_t$ and $\boldsymbol{P}^{GP,f}_t$ have been introduced in \eqref{eq:nGPprocess}, while $\boldsymbol{A}^{f^b}_t$ and $\boldsymbol{P}^{GP,f^b}_t$ have been introduced in \eqref{eq:forwardboundarymodel}.

Using Lemma \ref{lem:jointgaussian} two times, the joint mean $\boldsymbol{m}'$ of \eqref{eq:priorjoint} is
\begin{equation}
  \boldsymbol{m}' = \begin{pmatrix}
  \boldsymbol{A}_t \begin{pmatrix} \boldsymbol{A}^{f^b}_t \boldsymbol{m}_{t-1} \\
    \boldsymbol{m}_{t-1} \end{pmatrix} \\
  \boldsymbol{A}^{f^b}_t \boldsymbol{m}_{t-1} \\
  \boldsymbol{m}_{t-1}
  \end{pmatrix}
\end{equation}
and the covariance is
\begin{align}
  \tiny
  \boldsymbol{P}' = \begin{bmatrix}
    \boldsymbol{A}_t \tilde{\boldsymbol{P}}_{t-1} \boldsymbol{A}^T_t + \boldsymbol{P}^{GP,f}_t & \boldsymbol{A}_t \begin{bmatrix} \boldsymbol{S}^{f^b}_t \\ \boldsymbol{P}_{t-1} (\boldsymbol{A}^{f^b}_t)^T \end{bmatrix} & \boldsymbol{A}_t \begin{bmatrix} \boldsymbol{A}^{f^b}_t \boldsymbol{P}_{t-1} \\ \boldsymbol{P}_{t-1} \end{bmatrix} \\ 
    \left( \boldsymbol{A}_t \begin{bmatrix} \boldsymbol{S}^{f^b}_t \\ \boldsymbol{P}_{t-1} (\boldsymbol{A}^{f^b}_t)^T \end{bmatrix} \right)^T & \boldsymbol{S}^{f^b}_t & \boldsymbol{A}^{f^b}_t \boldsymbol{P}_{t-1} \\
    \left( \boldsymbol{A}_t \begin{bmatrix} \boldsymbol{A}^{f^b}_t \boldsymbol{P}_{t-1} \\ \boldsymbol{P}_{t-1} \end{bmatrix} \right)^T & \left( \boldsymbol{A}^{f^b}_t \boldsymbol{P}_{t-1} \right)^T & \boldsymbol{P}_{t-1}
  \end{bmatrix}.
\end{align}
Here we introduced 
\begin{align}
  \boldsymbol{S}^{f^b}_t {=}&\: \boldsymbol{A}^{f^b}_t \boldsymbol{P}_{t-1} \left( \boldsymbol{A}^{f^b}_t \right)^T + \boldsymbol{P}_t^{GP,f^b}, \\
  \tilde{\boldsymbol{P}}_{t-1} {=}&\: \begin{bmatrix} \boldsymbol{S}^{f^b}_t & \boldsymbol{A}^{f^b}_t \boldsymbol{P}_{t-1} \\
   \left( \boldsymbol{A}^{f^b}_t \boldsymbol{P}_{t-1} \right)^T & \boldsymbol{P}_{t-1} \end{bmatrix}.
\end{align}
The prior distribution 
\begin{equation}
  p(\boldsymbol{f}_t| \boldsymbol{f}^y_{1:t-1}, \boldsymbol{f}^b_{1:t}) = \mathrm{N} \left( \boldsymbol{f}_t | \boldsymbol{m}^-_t, \boldsymbol{P}^-_t \right),
\end{equation}
is now obtained from \eqref{eq:priorjoint} by first marginalizing over $\boldsymbol{f}_{t-1}$ and then conditioning on $\boldsymbol{f}^b_t$. 
Prior mean $\boldsymbol{m}^-_t$ and covariance $\boldsymbol{P}^-_t$ are
\begin{align}
  \boldsymbol{m}^-_t {=}&\: \boldsymbol{A}_t \begin{footnotesize} \begin{pmatrix} \boldsymbol{f}^b_t \\
  \boldsymbol{m}_{t-1} + \boldsymbol{P}_{t-1} \left( \boldsymbol{A}^{f^b}_t \right)^T \left( \boldsymbol{S}^{f^b}_t \right)^{-1} \left( \boldsymbol{f}^b_t - \boldsymbol{A}^{f^b}_t \boldsymbol{m}_{t-1} \right) \end{pmatrix} \end{footnotesize}, \\
  \boldsymbol{P}^-_t {=}&\: \boldsymbol{A}_t \tilde{\boldsymbol{P}}_{t-1} \boldsymbol{A}^T_t + \boldsymbol{P}^{GP,f}_t - \boldsymbol{A}_t \begin{bmatrix} \boldsymbol{S}^{f^b}_t \\ \boldsymbol{P}_{t-1} \left( \boldsymbol{A}^{f^b}_t \right)^T \end{bmatrix} \nonumber \\
  &{\times}\: \left( \boldsymbol{S}^{f^b}_t \right)^{-1} \begin{bmatrix} \boldsymbol{S}^{f^b}_t \\ \boldsymbol{P}_{t-1} \left( \boldsymbol{A}^{f^b}_t \right)^T \end{bmatrix}^T \boldsymbol{A}_t^T.
\end{align}
Equipped with this, we can calculate the joint distribution of the current state (prior distribution) and measurements (likelihood).
This will allow us to write down the posterior distribution later on.
Using Assumption 2, we have
\begin{align}
  {}&\:p(\boldsymbol{f}_t, \boldsymbol{f}^y_t | \boldsymbol{f}^y_{1:t-1},\boldsymbol{f}^b_{1:t}) \nonumber \\
  {=}&\: \underbrace{p(\boldsymbol{f}^y_t|\boldsymbol{f}_t)}_{\text{likelihood}} \underbrace{p(\boldsymbol{f}_t | \boldsymbol{f}^y_{1:t-1},\boldsymbol{f}^b_{1:t})}_{\text{prior}} \nonumber \\
  {=}&\: \mathrm{N} ( \boldsymbol{f}^y_t | \boldsymbol{C}_t \boldsymbol{f}_t, \boldsymbol{P}^{GP,f^y}_t ) \mathrm{N} \left(\boldsymbol{f}_t | \boldsymbol{m}^-_t, \boldsymbol{P}^-_t \right) \nonumber \\
  {=}&\: \mathrm{N} \left( \begin{pmatrix}
  \boldsymbol{f}_t \\
  \boldsymbol{f}^y_t \end{pmatrix} | \boldsymbol{m}'', \boldsymbol{P}'' \right).
  \label{eq:jointstatesandmeasurements}
\end{align}
The covariance matrices $\boldsymbol{C}_t$ and $\boldsymbol{P}^{GP,f^y}_t$ were introduced in \eqref{eq:nGPmeasurement}.
We use Lemma \ref{lem:jointgaussian} one more time to get the joint mean as
\begin{equation}
  \boldsymbol{m}'' = \begin{pmatrix}
    \boldsymbol{m}^-_t \\
    \boldsymbol{C}_t \boldsymbol{m}^-_t
    \end{pmatrix},
\end{equation}
and the covariance as
\begin{equation}
\boldsymbol{P}'' = \begin{bmatrix}
  \boldsymbol{P}^-_t & \boldsymbol{P}^-_t \boldsymbol{C}^T_t \\
  \boldsymbol{C}_t \boldsymbol{P}^-_t & \boldsymbol{C}_t \boldsymbol{P}^-_t \boldsymbol{C}^T_t + \boldsymbol{P}^{GP,f^y}_t
\end{bmatrix}.
\end{equation}
The posterior distribution
\begin{align}
p(\boldsymbol{f}_t | \boldsymbol{f}^y_t, \boldsymbol{f}^y_{1:t-1}, \boldsymbol{f}^b_{1:t}) \:{=}&\: p(\boldsymbol{f}_t | \boldsymbol{f}^y_{1:t}, \boldsymbol{f}^b _{1:t}) \nonumber \\
  {=}&\: \mathrm{N} \left( \boldsymbol{m}_t, \boldsymbol{P}_t \right)
\end{align}
is obtained by conditioning the joint distribution \eqref{eq:jointstatesandmeasurements} on the current measurement using Lemma \ref{lem:conditional}.
The posterior mean is
\begin{align}
 \boldsymbol{m}_t \:{=}&\: \boldsymbol{m}^-_t  + \boldsymbol{P}^-_t \boldsymbol{C}^T_t \nonumber \\
  &{\times}\: \left( \boldsymbol{C}_t \boldsymbol{P}^-_t \boldsymbol{C}^T_t  + \boldsymbol{P}^{GP,f^y}_t \right)^{-1} \left( \boldsymbol{f}^y_t - \boldsymbol{C}_t \boldsymbol{m}^-_t \right),
\end{align}
and the corresponding posterior variance is 
\begin{equation}
  \boldsymbol{P}_t = \boldsymbol{P}^-_t - \boldsymbol{P}^-_t \boldsymbol{C}^T_t \left( \boldsymbol{C}_t \boldsymbol{P}^-_t \boldsymbol{C}^T_t + \boldsymbol{P}^{GP,f^y}_t \right)^{-1} \boldsymbol{C}_t \boldsymbol{P}^-_t.
\end{equation}
To summarize the numerical GPKF, we have the prediction step as
\begin{align}
  \boldsymbol{m}^-_t {=}&\: \boldsymbol{A}_t \begin{footnotesize} \begin{pmatrix} \boldsymbol{f}^b_t \\
  \boldsymbol{m}_{t-1} + \boldsymbol{P}_{t-1} \left( \boldsymbol{A}^{f^b}_t \right)^T \left( \boldsymbol{S}^{f^b}_t \right)^{-1} \left( \boldsymbol{f}^b_t - \boldsymbol{A}^{f^b}_t \boldsymbol{m}_{t-1} \right) \end{pmatrix} \end{footnotesize}, \\
  \boldsymbol{P}^-_t {=}&\: \boldsymbol{A}_t \tilde{\boldsymbol{P}}_{t-1} \boldsymbol{A}^T_t + \boldsymbol{P}^{GP,f}_t - \boldsymbol{A}_t \begin{bmatrix} \boldsymbol{S}^{f^b}_t \\ \boldsymbol{P}_{t-1} \left( \boldsymbol{A}^{f^b}_t \right)^T \end{bmatrix} \nonumber \\
  &{\times}\: \left( \boldsymbol{S}^{f^b}_t \right)^{-1} \begin{bmatrix} \boldsymbol{S}^{f^b}_t \\ \boldsymbol{P}_{t-1} \left( \boldsymbol{A}^{f^b}_t \right)^T \end{bmatrix}^T \boldsymbol{A}_t^T.
  \label{eq:KF_predictionstep_summary}
\end{align}
and the update step as
\begin{align}
\boldsymbol{v}_t &= \boldsymbol{f}^y_t - \boldsymbol{C}_t \boldsymbol{m}^-_t, \\
\boldsymbol{S}_t &= \boldsymbol{C}_t \boldsymbol{P}_t^- \boldsymbol{C}^T_t + \boldsymbol{P}^{GP,f^y}_t, \\
\boldsymbol{K}_t &= \boldsymbol{P}^-_t \boldsymbol{C}^T_t (\boldsymbol{S}_t)^{-1}, \\
\boldsymbol{m}_t &= \boldsymbol{m}^-_t + \boldsymbol{K}_t \boldsymbol{v}_t, \\
\boldsymbol{P}_t &= \boldsymbol{P}_t^- - \boldsymbol{K}_t \boldsymbol{S}_t (\boldsymbol{K}_t)^T.
\label{eq:KF_updatestep_summary}
\end{align}
Due to having to include boundary conditions, the prediction step doesn't perfectly align with the traditional KF.
It is not possible to simply write the prior mean prediction as the previous posterior mean mapped through the dynamic matrix $\boldsymbol{A}_t$.
Treating boundary conditions as measurements results in much simpler expressions that smoothly align with the traditional KF, see \cite{kueper2020ifacnGPKF}.
Of course this means that prediction steps do not abide boundary conditions which can be problematic when multiple prediction steps have to be computed between update steps.

\subsection*{Hyper-parameter estimation via the marginal likelihood}
\label{subsec:NLML_KF}
Hyper-parameters are estimated in each update step.
In GP regression the hyper-parameters are usually estimated by minimizing the negative log marginal likelihood (NLML).
The marginal likelihood of the Kalman filter, marginal with respect to the current state $f_t$, is found in the recursive Bayes denominator \eqref{eq:numericalGPKFBayes} as
\begin{align}
	p(\boldsymbol{f}^y_t|\boldsymbol{f}^y_{1:t-1}, \boldsymbol{f}^b_{1:t}) = \mathrm{N} \left( \boldsymbol{C}_t \boldsymbol{m}^-_t,\, \boldsymbol{S}_t \right).
	\label{eq:MLGPKF}
\end{align}
The corresponding NLML is
\begin{align}
	{}&{-}\: \log p(\boldsymbol{f}^y_t|\boldsymbol{f}^y_{1:t-1}, \boldsymbol{f}^b_{1:t}) \\
  {=}&\: \dfrac{1}{2} \left( \boldsymbol{f}^y_t - \boldsymbol{C}_t \boldsymbol{m}^-_t \right)^T \boldsymbol{S}^{-1}_t \left( \boldsymbol{f}^y_t - \boldsymbol{C}_t \boldsymbol{m}^-_t \right) \nonumber \\
	{}&{+}\: \dfrac{1}{2} \log (\det(\boldsymbol{S}_t)) + \dfrac{N_y}{2} \log (2 \pi),
  \label{eq:NLMLKF} 
\end{align}
with $N_y$ being the number of measurements.
The partial derivatives of this NLML required for minimization can be calculated analytically and are shown in Appendix \ref{app:NLMLderivation}.
\section{Simulation case studies}
\label{sec:casestudies}

In this section we will showcase the numerical Gaussian process Kalman filter (GPKF) using different case studies.
For all case studies the relative error between posterior mean estimate $\boldsymbol{m}_t$ and analytical solution $\boldsymbol{f}_{t,ref}$ evaluated on the same points $\boldsymbol{X}$ is defined as
\begin{align}
\text{relative error at time} \: t &= \dfrac{\lVert \boldsymbol{f}_{t,ref} - \boldsymbol{m}_t \rVert}{\lVert \boldsymbol{f}_{t,ref} \rVert},
\label{eq:rel_error}
\end{align}
with euclidean distance $\lVert \cdot \rVert$.
All case studies have been implemented in MATLAB.

\subsection{Step shaped one-dimensional advection equation}
\label{subsec:onedim_casestudy}
We study the one-dimensional advection equation
\begin{align}
	\dfrac{\partial }{\partial t} f(t,x) &= -g \dfrac{\partial f}{\partial x}, 
	\label{eq:step_1dadvection_casestudy}
\end{align}
with initial and boundary condition as
\begin{align}
	f(0,x) &= 2H(x) - 1, \\
	f(t,0) &= 0.
\end{align}
Here, $H(x)$ is the Heaviside step function.
We solve \eqref{eq:step_1dadvection_casestudy} with the implicit Euler and a step size of $\Delta t = 0.005$.
The GP prior is $f_t \sim \text{GP} (0, k^{ff}_{t,t}(x,x'))$ with a neural network covariance function
\begin{align}
	k^{ff}_{t,t}(x,x') = \tfrac{2}{\pi} \sin^{-1} \left( \tfrac{2 (\sigma_0^2 + \sigma^2 x x')}{\sqrt{( 1 + 2(\sigma_0^2 + \sigma^2 x^2) ) ( 1 + 2(\sigma_0^2 + \sigma^2 x'^2) )}} \right).
\end{align}
This covariance function is capable of capturing discontinuities.
The hyper-parameters are $\sigma_0^2$ and $\sigma^2$.
Measurements are noisy point evaluations so that the measurement equation reads as
\begin{align}
	f^y_t (y) = f_t(x) + r_t,
\end{align}
with $r_t \sim \text{N} (0, \sigma^2_{r,true})$ being white noise.
Measurements are received every third model step size so that between each KF update step there are three KF prediction steps.
Hyper-parameters are learned by minimizing the NLML \eqref{eq:NLMLKF} in each update step.

\begin{figure*}[tb!]
\centering
\begin{subfigure}[t]{0.48\textwidth}
	\includegraphics[width=0.99\columnwidth]{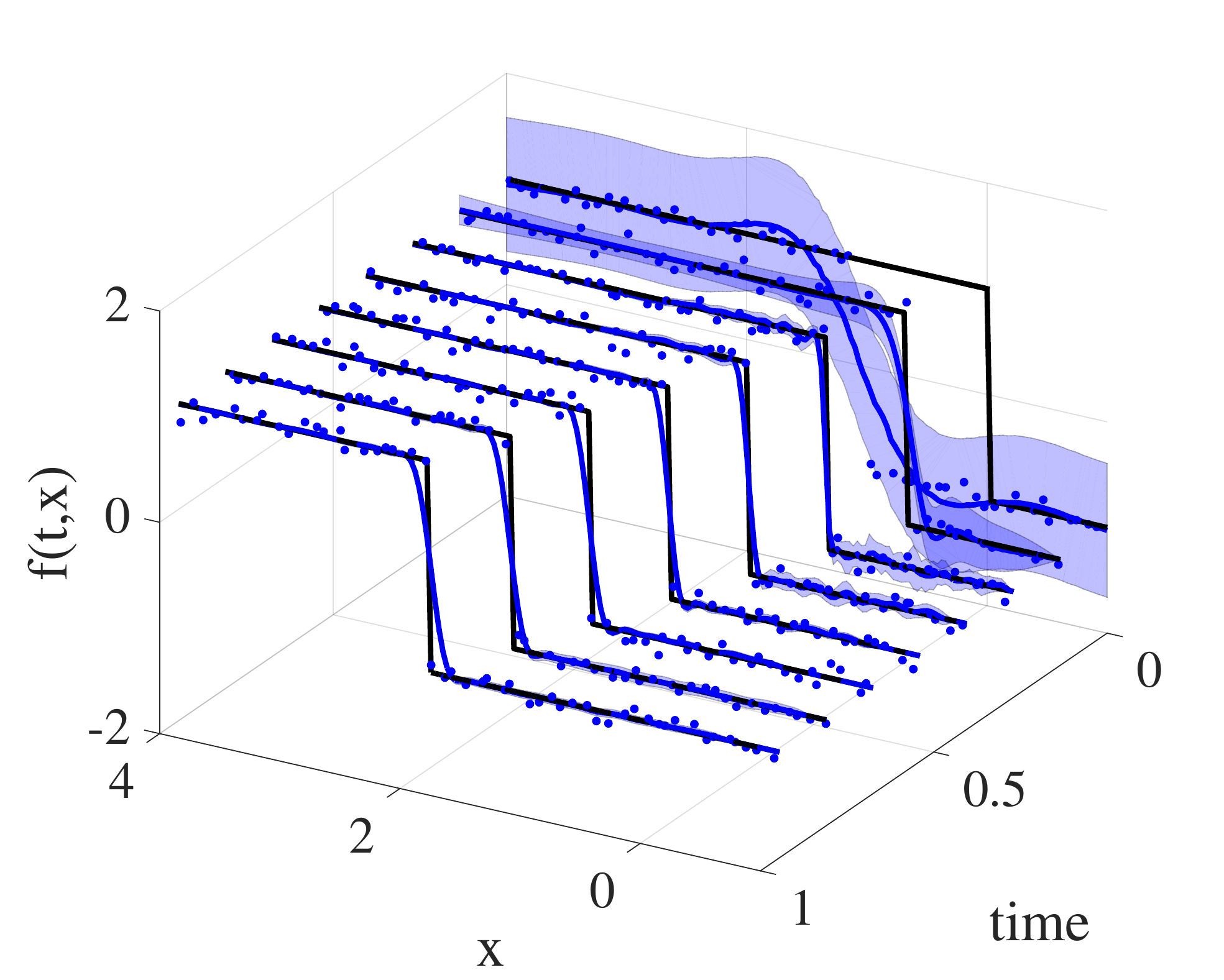}
	\caption{Snapshots for the step shaped one-dimensional advection equation.
	Noisy measurements are blue dots, while the true solution is represented by black lines.
	Posterior mean estimates along with their $95\,\%$ confidence intervals are shown in blue.}
	\label{subfig:snapshots_1d_step}
\end{subfigure} \hspace*{0.25cm}
\begin{subfigure}[t]{0.48\textwidth}
	\centering
	\includegraphics[width=0.99\columnwidth]{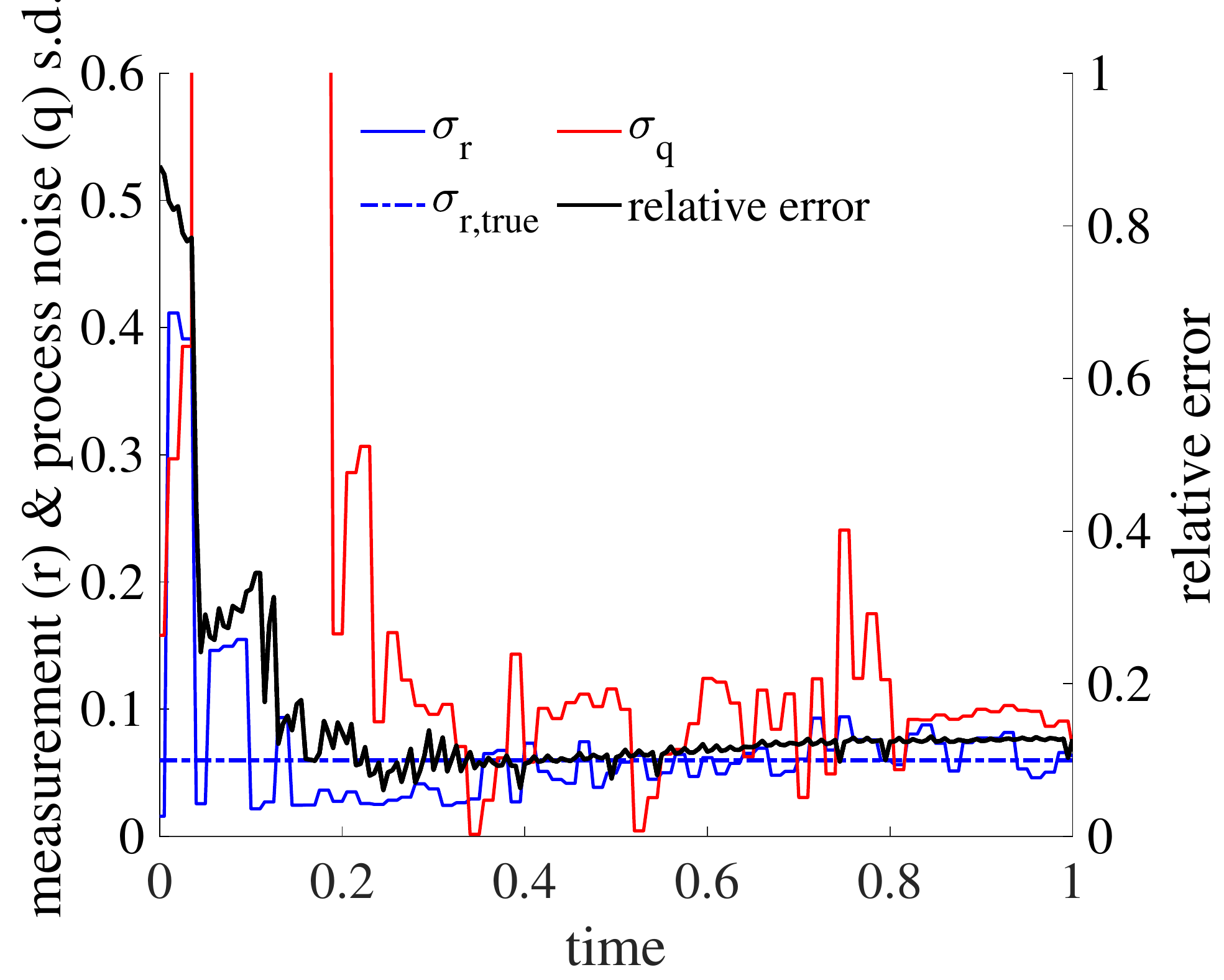}
	\caption{Hyper-parameter estimation and relative error \eqref{eq:rel_error} over time.
	Estimated process noise $\sigma_q$ gets exceedingly large at the beginning.
	This is cut off from the plot.}
	\label{subfig:hyper_1d_step}
\end{subfigure}
\caption{Estimation results for the step shaped one-dimensional advection equation \eqref{eq:step_1dadvection_casestudy}.}
\label{fig:advection_1d_step}
\end{figure*}

Estimation results are illustrated in Fig.~\ref{fig:advection_1d_step}.
The initial estimate (blue curve with uncertainty band in Fig. \ref{subfig:snapshots_1d_step}) is far off from the true solution (black line).
However, rather quickly the estimate converges closer, as indicated by a dropping relative error (black curve in Fig. \ref{subfig:hyper_1d_step}).
The relative error drops to a value of around $0.1$ before it slowly starts to increase.
This divergence is not uncommon in Kalman filtering \cite{fitzgerald1971divergence}.
One possible cause, that is also present in Fig.~\ref{subfig:snapshots_1d_step}, is a low posterior covariance matrix $\boldsymbol{P}_t$.
As a result, the KF puts less importance on measurements and more on model predictions, which in this case aren't perfect due to the temporal discretization required for numerical GPs.
Remedies such as the Kalman filter with fading memory exist \cite{simon2006optimal}.

The measurement noise level $\sigma_r$ is correctly learned online to be $\sigma_{r,true}$.
Estimated process noise $\sigma_q$ is non-zero, likely reflecting the aforementioned model imperfection.
For a few time steps at the beginning its estimation gets exceedingly large.
This is cut off from the plot.

\subsection{Liouville's equation}
\label{subsec:harmonicoscillator}
\begin{figure}[tb]
	\centering
	\includegraphics[width=1\columnwidth]{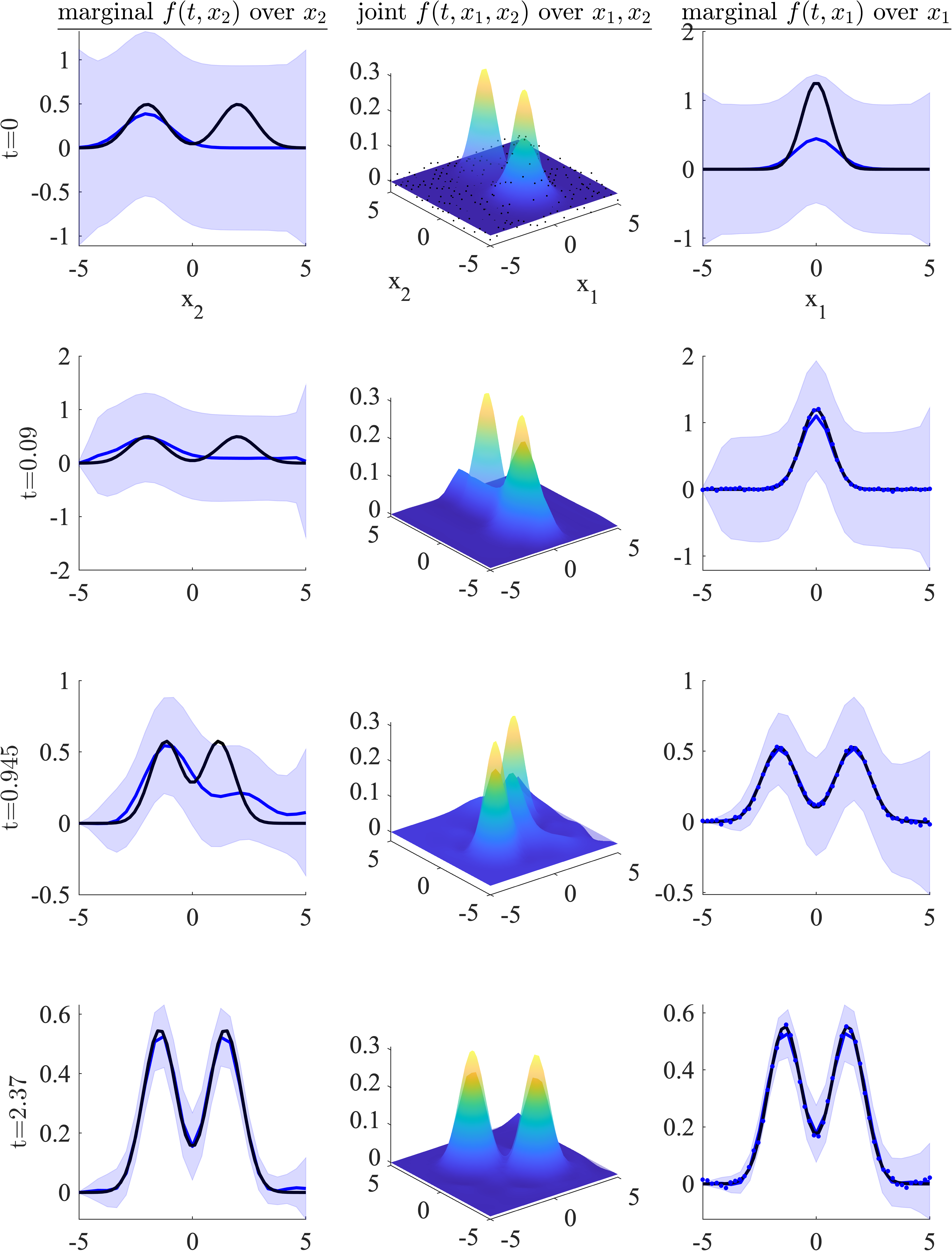}
	\caption{Selected snapshot plots of Liouville's equation.
	Left and right plots show the marginal distributions $f(t,x_2)$ and $f(t,x_1)$, respectively.
	Posterior mean estimates of the marginals are shown in blue along with their $95\,\%$ confidence intervals (CI), while black curves show the analytical solution.
	Right plots also show noisy measurements of $f(t,x_1)$ as blue circles.
	Middle plots show the posterior mean estimate for the joint distribution $f(t,x_1,x_2)$, as well as the analytical solution (transparent).
	Uncertainty bands have been omitted. 
	The initial estimate is based on the black dots (first row, middle plot).
	Online measurements of $f(t,x_1)$ shown as blue dots (right plots).}
	\label{fig:snapshots_2D_and_marginals}
\end{figure}
We study Liouville's equation
\begin{align}
\dfrac{\partial}{\partial t} f(t,\boldsymbol{x}) &= - \mathrm{div} \left( f(t,\boldsymbol{x}) \dot{\boldsymbol{x}} \right) \\
&= - \dot{x}_1 \dfrac{\partial f}{\partial x_1} - f \dfrac{\partial \dot{x}_1}{\partial x_1} - \dot{x}_2 \dfrac{\partial f}{\partial x_2} - f \dfrac{\partial \dot{x}_2}{\partial x_2},
\label{eq:2dliouvilles}
\end{align}
wherein the vector field $\dot{\boldsymbol{x}}$ is governed by
\begin{align}
\dot{\boldsymbol{x}} &=
\begin{bmatrix}
0 & 1 \\
-1 & 0
\end{bmatrix}
\boldsymbol{x}.
\end{align}
Initial and boundary conditions are
\begin{align}
	f(0,x_1,x_2) &= \text{N} \left( \boldsymbol{\mu}_1, \Sigma \right) + \text{N} \left( \boldsymbol{\mu}_2, \Sigma \right), \\
	f(t,0,x_2) &= 0, \\
	f(t,x_1,0) &= 0.
\end{align}
Here, the initial condition consists of two Gaussian bumps sitting opposite of each other at $\boldsymbol{\mu}_1 = (0, -2)^T$ and $\boldsymbol{\mu}_2 = (0, 2)^T$ with covariance matrix $\Sigma = \text{diag}(0.4, 0.65)$.
The initial estimate communicated to the Kalman filter consists of only one Gaussian bump located at $\boldsymbol{\mu}_1$ with increased variance $\Sigma + \text{diag}(0.4, 0.4)$.

We discretize \eqref{eq:2dliouvilles} in time using the implicit Euler method with a step size of $\Delta t = 0.005$, and place a GP prior on $f_t\sim \mathrm{GP}(0, k^{ff}_{t,t}(x,x'))$ with the following squared exponential covariance function
\begin{align}
	k^{ff}_{t,t}(x,x') &= \sigma_n^2 \exp \left( - \dfrac{(x_1 - x'_1)^2}{2 l_1^2} - \dfrac{(x_2 - x'_2)^2}{2 l_2^2} \right).
\end{align}
Here, the hyper-parameters are $\sigma_n^2$, $l_1$, and $l_2$.
Through noisy measurements we receive the marginal distribution of $x_1$ as
\begin{align}
	f_t^y(y=x_1) &= \int_{x_{2,min}}^{x_{2,max}} f_t(x_1,x_2) \text{d} x_2 + r_t,
\end{align} 
with $r_t \sim \text{N} (0, \sigma^2_{r,true})$ being white noise.

The vector field $\dot{\boldsymbol{x}}$ describes harmonic oscillations, plotting the solution $f(t, \boldsymbol{x})$ results therefore in circular motions.
The simulation runs for one full rotation $T_{sim} = 2 \pi$.
Although measurements are formally the $x_1$-marginal, over time measurements will be marginals of the initial distribution from all angles due to the circular motion of the joint distribution $f(t,\boldsymbol{x})$.
Interested readers are referred to \cite{zeng2019sample} for further details.
As discovered and explained by \cite{zeng2015ensemble}, this case study illustrates the relatedness to tomography.
Measurements are received every $\tfrac{T_{sim}}{64}$, which translates to 19 KF predictions steps for every KF update step.

\begin{figure}[tb]
\begin{center}
	\includegraphics[width=0.9\columnwidth]{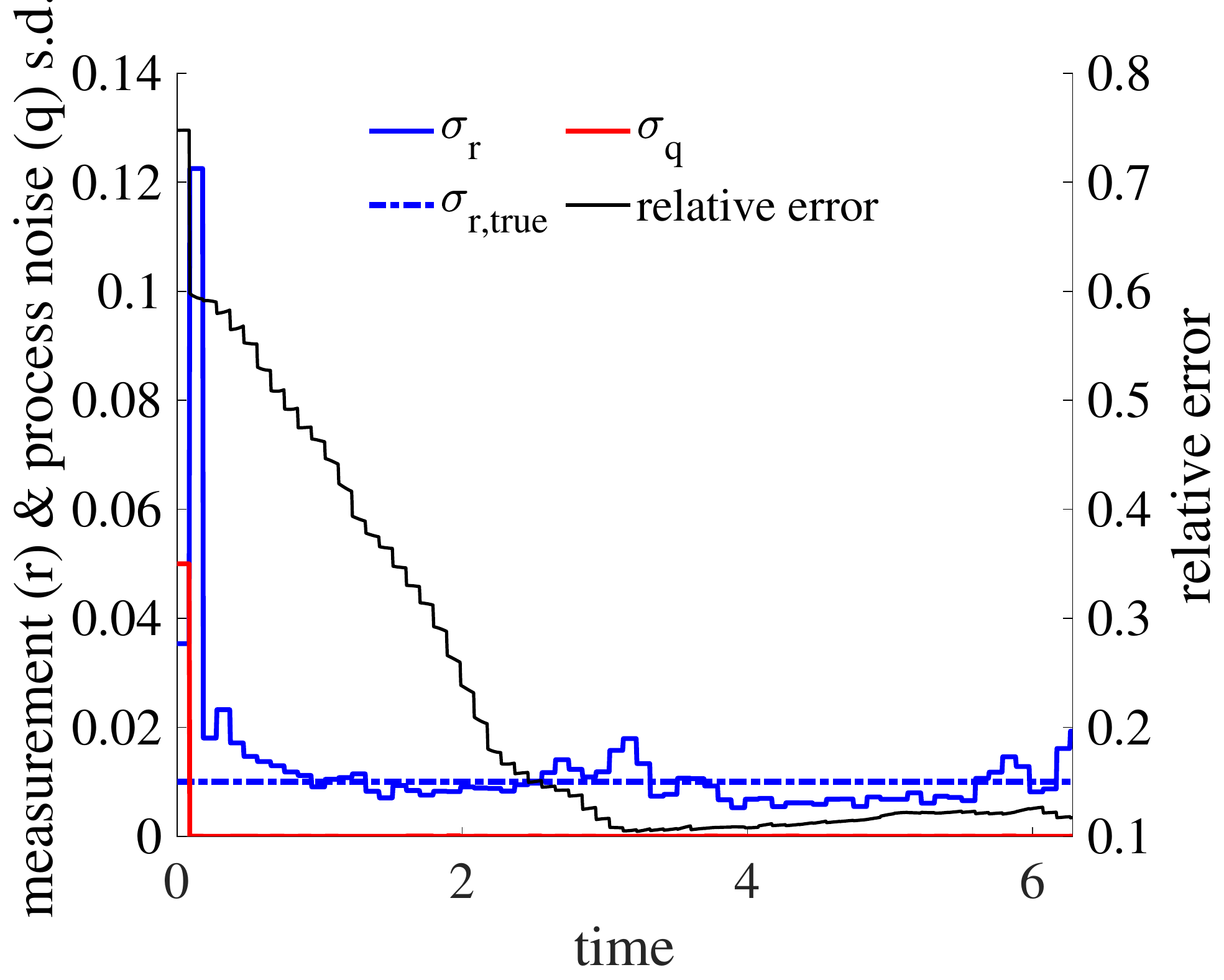}
	\caption{Noise hyper-parameter estimation and relative estimation error for Liouville's equation.}
	\label{fig:hyper_rel_error_2doscillations}
\end{center}
\end{figure}

State estimation results are displayed in Fig.~\ref{fig:snapshots_2D_and_marginals} while noise level estimations are displayed in Fig.~\ref{fig:hyper_rel_error_2doscillations}.
Through measurements of the $x_1-$marginal $f(t,x_1)$ the numerical GPKF is able to reconstruct the latent $x_2-$marginal distribution $f(t,x_2)$ and the joint distribution $f(t,x_1,x_2)$.
Noise level estimation $\sigma_r$ oscillates around its true level $\sigma_{r,true}$, while process noise $\sigma_q$ is set to zero by the numerical GPKF.
The relative estimation error drops significantly before settling in at around $0.16$.

Hyper-parameter optimization requires an ad-hoc approach for this case study.
Simply using the marginal likelihood $p(\boldsymbol{f}^y_t|\boldsymbol{f}^y_{1:t-1}, \boldsymbol{f}^b_{1:t})$ of the Kalman filter as before will result in ever larger hyper-parameter values of the marginalized dimension, i.e. $l_2$.
We conjecture that due to the marginalization of $x_2$, its impact on the marginal likelihood has been largely nullified, rendering $l_2$ unidentifiable.
A remedy to this would be to use the Radon-transform and thereby treat this case study as a problem of tomographic type. 
Through a change of variables, known as input warping in the GP community, the Radon-transform would convolute the dimensions $x_1$ and $x_2$, and therefore also both length-scales $l_1$ and $l_2$, thereby preventing cancellation of $l_2$ through the measurement marginalization.
See \cite{purisha2019probabilistic} for the use of GPs in tomographic inverse problems.

Instead, we successfully employed the function 
\begin{align}
	{}&{-}\: \log p(\boldsymbol{f}^y_t, \boldsymbol{f}^b_t, \boldsymbol{f}_{t-1}) \nonumber \\
	{=}&{}\:\dfrac{1}{2} \begin{pmatrix} \boldsymbol{f}^y_t \\ \boldsymbol{f}^b_t \\ \boldsymbol{m}_{t-1} \end{pmatrix}^T 
	\underbrace{\begin{bmatrix} \boldsymbol{K}^{f^y f^y}_{t,t} & \boldsymbol{K}^{f^y f^b}_{t,t} & \boldsymbol{K}^{ff}_{t,t-1} \\
	 \boldsymbol{K}^{f^b f^y}_{t,t} & \boldsymbol{K}^{f^b f^b}_{t,t} & \boldsymbol{K}^{f^b f}_{t,t-1} \\
	 \boldsymbol{K}^{ff^y}_{t-1,t} & \boldsymbol{K}^{ff^b}_{t-1,t} & \boldsymbol{K}^{ff}_{t-1,t-1}
	\end{bmatrix}^{-1}}_{=\boldsymbol{K}}
	\begin{pmatrix} \boldsymbol{f}^y_t \\ \boldsymbol{f}^b_t \\ \boldsymbol{m}_{t-1}\end{pmatrix} \nonumber \\
	{}&{+}\: \dfrac{1}{2} \log (\det(\boldsymbol{K})) + \dfrac{N}{2} \log (2 \pi)
\end{align}
as the objective function to minimize.
Here, we used the previous posterior mean $\boldsymbol{m}_{t-1}$ as fixed targets for $\boldsymbol{f}_{t-1}$.
The number of data points is $N$.
This objective function now contains both length-scales $l_1$ and $l_2$ explicitly and therefore works.
\section{Discussion}
\label{sec:discussion}

In this section we elaborate on computational load, optimality, observability, and distinguishing features of the numerical Gaussian process Kalman filter.

\subsection*{On computational cost}
\label{subsec:computationalcost}
The main computational cost lies in the hyper-parameter optimization which requires the inversion of the measurement covariance matrix.
This is inherent to GPs and scales cubically with the number of training points.
There are however different approximations to reduce this to less than cubic scaling, see e.g. \cite{snelson2006sparse,solin2020hilbert}.

Actual simulation of the model with numerical GPs, i.e. propagation through time, does not require numerical integration and instead only involves algebraic manipulations as given by \eqref{eq:condmeanandcov}.
The two dimensional simulation ran for $111\,$min, while the one dimensional simulation ran for $9.5\,$min on a 2020 $13"$ MacBook Pro $2,3\,\text{GHz}$ quad-core i7.

In addition to the number of dimensions $\boldsymbol{x} \in \mathbb{R}^{d_x}$, computational load is also influenced by the prior covariance function and the PDE structure.
Both can lead to less or more difficult to handle contours of the NLML, thereby influencing the computational load via the hyper-parameter optimization.
For example, we found simulations of the same PDE using a squared exponential covariance function to run faster and more robustly than using a neural network covariance function.

\subsection*{On optimality}
\label{subsec:optimality}
For a linear and Gaussian system, the Kalman filter gives the optimal estimate of the state $\boldsymbol{m}_t^{opt}$ given the measurement history $\boldsymbol{n}^y_{1:t}$.
For spatiotemporal systems, we also condition on the boundary history $\boldsymbol{n}^b_{1:t}$.
Optimal is defined in this context as
\begin{align}
{}&{}\:\boldsymbol{m}_t^{opt}(\boldsymbol{n}^y_{1:t},\boldsymbol{n}^b_{1:t},\boldsymbol{\theta}) \nonumber \\
{=}&{}\: \arg\,\min_{\boldsymbol{m}_t} \text{E} \left[ (\boldsymbol{n}_t - \boldsymbol{m}_t)^T (\boldsymbol{n}_t - \boldsymbol{m}_t) | \boldsymbol{n}^y_{1:t}, \boldsymbol{n}^b_{1:t}, \boldsymbol{\theta} \right].
\end{align}
Here, the estimated state is $\boldsymbol{m}_t$ and the true (stochastic) state is $\boldsymbol{n}_t$.
The hyper-parameters of the Gaussian process are denoted as $\boldsymbol{\theta}$.
They influence the state estimate through i) the covariance function itself and ii) their estimated values.

Although working with spatiotemporal systems, the input space of numerical GPs is solely the spatial domain.
Choosing an appropriate covariance function should therefore be based upon the solution shape with respect to spatial dimensions.
For the step shaped one-dimensional advection equation for example, we chose the neural network covariance function due to its ability to handle discontinuities.
In the original numerical GP work \cite{raissi2018numerical} the Burgers equation is solved with the neural network covariance function as well because its solution develops a shock front.

As showcased and argued in \cite{williams2006gaussian}, multiple optima of the marginal likelihood can certainly exist, but are usually not a problem.
Indeed, the different optima represent different interpretations of the data.

\subsection*{On observability}
\label{subsec:obs}

We showed that numerical Gaussian processes can be brought into probabilistic state space model form.
By design of numerical GPs, this state space model is structured by the underlying time-dependent partial differential equation.
Showing observability of the PDE might therefore be transferable to the numerical GP state space model \eqref{eq:nGPprocess}, \eqref{eq:nGPmeasurement}.

Closely connected to observability is the concept of identifiability.
This relates to the estimation of (hyper-)parameters.
As seen and explained in the second case study involving Liouville's equation, an ad-hoc replacement for the marginal likelihood might be necessary to make hyper-parameter estimation work.

\subsection*{On distinguishing features}
\label{subsec:}
Kalman filters require fine-tuning before they deliver acceptable performance.
This fine-tuning involves adjusting the process $\boldsymbol{Q}$ and measurement noise covariances $\boldsymbol{R}$, as well as the initial error covariance matrix $\boldsymbol{P}_0$.

Fine-tuning of numerical GPKFs is either much less tedious or not required at all.
Once initialized with reasonable values, process and measurement noise level hyper-parameters $\sigma_q$ and $\sigma_r$ are adapted online to the measurement stream. 

As for the initial error covariance matrix $\boldsymbol{P}_0$, no manual fine-tuning was required for the case studies due to the probabilistic nature of GPs.
$\boldsymbol{P}_0$ was simply obtained by calculating the posterior GP variance \eqref{eq:condmeanandcov} for given initial data.
This was done without optimizing the hyper-parameters, as can be seen by the bad initial fit to the step function in Fig.~\ref{subfig:snapshots_1d_step}.

\section{Conclusions}
\label{sec:conclusions}

In this article, we presented a new type of Kalman filter for spatiotemporal systems.
Building on numerical Gaussian processes, we developed a probabilistic state space model that is linear and has Gaussian distributed states.
These properties made it possible to derive the Kalman filter algorithm based on numerical Gaussian process state space models.

The resulting numerical Gaussian process Kalman filter was showcased with a step shaped one-dimensional advection equation where measurements were noisy point evaluations of the solution.
In another case study Liouville's equation with integral measurements was studied.
In both case studies the numerical GPKF was able to reconstruct the latent state and also estimate the true measurement noise levels through its hyper-parameters.

\appendix

Lemmata \ref{lem:jointgaussian} and \ref{lem:conditional} are taken from \cite{sarkka2013bayesian}.
\subsection{Joint distribution of Gaussian variables}
\label{lem:jointgaussian}
If random variables $\boldsymbol{x} \in \mathbb{R}^{d_x}$ and $\boldsymbol{y} \in \mathbb{R}^{d_y}$ have the Gaussian probability distributions
\begin{align}
  \boldsymbol{x} &\sim \mathrm{N} \left(\boldsymbol{m}, \boldsymbol{P} \right) \\
  \boldsymbol{y|x} &\sim \mathrm{N} \left(\boldsymbol{H}\boldsymbol{x} + \boldsymbol{u}, \boldsymbol{R} \right),
\end{align}
than the joint distribution of $\boldsymbol{x},\,\boldsymbol{y}$ and the marginal distribution of $\boldsymbol{y}$ are given as
\begin{align}
  \begin{pmatrix}
    \boldsymbol{x} \\
    \boldsymbol{y}
    \end{pmatrix} &\sim \mathrm{N} \left(
    \begin{pmatrix}
      \boldsymbol{m} \\
      \boldsymbol{H} \boldsymbol{m} + \boldsymbol{u}
    \end{pmatrix}, 
    \begin{pmatrix}
      \boldsymbol{P} & \boldsymbol{P} \boldsymbol{H}^T \\
      \boldsymbol{H}\boldsymbol{P} & \boldsymbol{H}\boldsymbol{P}\boldsymbol{H}^T + \boldsymbol{R}
    \end{pmatrix}
    \right), \\
    \boldsymbol{y} &\sim \mathrm{N} \left( \boldsymbol{H}\boldsymbol{m} + \boldsymbol{u}, \boldsymbol{H}\boldsymbol{P}\boldsymbol{H}^T + \boldsymbol{R} \right).
\end{align}
%
\subsection{Conditional distribution of Gaussian variables}
\label{lem:conditional}
If the random variables $\boldsymbol{x}$ and $\boldsymbol{y}$ have the joint Gaussian probability distribution
\begin{align}
  \begin{pmatrix}
    \boldsymbol{x} \\
    \boldsymbol{y}
    \end{pmatrix} &\sim \mathrm{N} \left(
    \begin{pmatrix}
      \boldsymbol{a} \\
      \boldsymbol{b}
    \end{pmatrix}, 
    \begin{pmatrix}
      \boldsymbol{A} & \boldsymbol{C} \\
      \boldsymbol{C}^T & \boldsymbol{B}
    \end{pmatrix}
    \right),
\end{align}
than the conditional distribution is
\begin{equation}
  \boldsymbol{x}|\boldsymbol{y} \sim \mathrm{N} \left( \boldsymbol{a} + \boldsymbol{C}\boldsymbol{B}^{-1}(\boldsymbol{y}-\boldsymbol{b}), \boldsymbol{A} - \boldsymbol{C}\boldsymbol{B}^{-1}\boldsymbol{C}^T \right).
\end{equation}

\subsection{Numerical Gaussian process state space model precursor}
\label{app:nGPSSprecursor}
We can derive the covariance functions of the joint model \eqref{eq:multioutputGPfull}, here shown using the implicit Euler, as
\begin{align}
\boldsymbol{K}_{GP} \:{=}&\: 
\mathrm{E} \left[
\begin{pmatrix}
f_t(x) \\
f^y_t(y) \\
f^bt(x_b) \\
f_{t-1}(x)
\end{pmatrix}
\begin{pmatrix}
f_t(x') \\
f^y_t(y') \\
f^b_t(x_b') \\
f_{t-1}(x')
\end{pmatrix}^T
\right] \\
\:{=}&\: \mathrm{E} \left[
\begin{pmatrix}
f_t \\
\mathcal{H}_x f_t + r_t \\
\mathcal{B}_x f_t \\
\mathcal{A}_x f_{t} - \Delta t q_{t-1}
\end{pmatrix}
\begin{pmatrix}
f_t \\
\mathcal{H}_{x'} f_t + r_t \\
\mathcal{B}_{x'} f_t \\
\mathcal{A}_{x'} f_{t} - \Delta t q_{t-1}
\end{pmatrix}^T
\right] \nonumber \\
\:{=}&\: \begin{footnotesize} \begin{bmatrix}
k & \mathcal{H}_{x'} k & \mathcal{B}_{x'} k & \mathcal{A}_{x'} k \\
\mathcal{H}_{x} k & \mathcal{H}_{x} \mathcal{H}_{x'} k + k^{rr}_{t,t} & \mathcal{H}_x \mathcal{B}_{x'} k & \mathcal{H}_x \mathcal{A}_{x'} k \\
\mathcal{B}_x k & \mathcal{B}_x \mathcal{H}_{x'} k & \mathcal{B}_x \mathcal{B}_{x'} k & \mathcal{B}_x \mathcal{A}_{x'} k \\
\mathcal{A}_x k & \mathcal{A}_x \mathcal{H}_{x'} k & \mathcal{A}_x \mathcal{B}_{x'} k & \mathcal{A}_x \mathcal{A}_{x'} k + \Delta t^2 k_{t,t}^{qq} 
\end{bmatrix} \end{footnotesize}.
\end{align}
Here, the prior covariance function $k^{ff}_{t,t}$ has been abbreviated as $k$.

\subsection{Partial derivatives of the KF NLML}
\label{app:NLMLderivation}
We had
\begin{align}
  &{-}\: \dfrac{\partial }{\partial \theta_j} \log p(\boldsymbol{f}^y_t|\boldsymbol{f}^y_{1:t-1}, \boldsymbol{f}^b_{1:t}) \nonumber \\
  {=}&\: \dfrac{1}{2} \dfrac{\partial }{\partial \theta_j} \left( \boldsymbol{f}^y_t - \boldsymbol{C}_t \boldsymbol{m}^-_t \right)^T \boldsymbol{S}^{-1}_t \left( \boldsymbol{f}^y_t - \boldsymbol{C}_t \boldsymbol{m}^-_t \right) \nonumber \\
  &{+}\: \dfrac{1}{2} \dfrac{\partial }{\partial \theta_j} \log (\det(\boldsymbol{S}_t)). 
  \label{eq:NLML_KF_derivative_app}
\end{align}
Expanding the first term and then using the product rule on it to evaluate all necessary partial derivatives gives
\begin{align}
&\: \dfrac{\partial }{\partial \theta_j} \left( \boldsymbol{f}^y_t - \boldsymbol{C}_t \boldsymbol{m}^-_t \right)^T \boldsymbol{S}^{-1}_t \left( \boldsymbol{f}^y_t - \boldsymbol{C}_t \boldsymbol{m}^-_t \right) \nonumber \\
{=}&\: \dfrac{\partial }{\partial \theta_j} \bigg( (\boldsymbol{n}^y_t)^T \boldsymbol{S}^{-1}_t \boldsymbol{f}^y_t - (\boldsymbol{f}^y_t)^T \boldsymbol{S}^{-1}_t \boldsymbol{C}_t \boldsymbol{m}^-_t \nonumber \\
&{-}\: (\boldsymbol{C}_t \boldsymbol{m}^-_t)^T \boldsymbol{S}^{-1}_t \boldsymbol{f}^y_t + (\boldsymbol{C}_t \boldsymbol{m}^-_t)^T \boldsymbol{S}^{-1}_t \boldsymbol{C}_t \boldsymbol{m}^-_t \bigg) \nonumber \\
{=}&\: (\boldsymbol{f}^y_t)^T \dfrac{\partial \boldsymbol{S}^{-1}_t}{\partial \theta_j} \boldsymbol{f}^y_t \nonumber \\
&{-}\: (\boldsymbol{f}^y_t)^T \left( \dfrac{\partial \boldsymbol{S}^{-1}_t}{\partial \theta_j} \boldsymbol{C}_t \boldsymbol{m}^-_t + \boldsymbol{S}^{-1}_t \dfrac{\partial \boldsymbol{C}_t}{\partial \theta_j} \boldsymbol{m}^-_t + \boldsymbol{S}^{-1}_t \boldsymbol{C}_t \dfrac{\partial \boldsymbol{m}^-_t}{\partial \theta_j} \right) \nonumber \\
&{-}\: \left( \dfrac{\partial \left(\boldsymbol{m}^-_t\right)^T}{\partial \theta_j} \boldsymbol{C}_t^T \boldsymbol{S}^{-1}_t + (\boldsymbol{m}^-_t)^T \dfrac{\partial \boldsymbol{C}_t^T}{\partial \theta_j} \boldsymbol{S}^{-1}_t \right) \boldsymbol{f}^y_t \nonumber \\
&{-}\: (\boldsymbol{m}^-_t)^T \boldsymbol{C}_t^T \dfrac{\partial \boldsymbol{S}^{-1}_t}{\partial \theta_j} \boldsymbol{f}^y_t \nonumber \\
&{+}\: \dfrac{\partial (\boldsymbol{m}^-_t)^T}{\partial \theta_j} \boldsymbol{C}_t^T \boldsymbol{S}^{-1}_t \boldsymbol{C}_t \boldsymbol{m}^-_t + (\boldsymbol{m}^-_t)^T \dfrac{\partial \boldsymbol{C}_t^T}{\partial \theta_j} \boldsymbol{S}^{-1}_t \boldsymbol{C}_t \boldsymbol{m}^-_t \nonumber \\
&{+}\: (\boldsymbol{m}^-_t)^T \boldsymbol{C}_t^T \dfrac{\partial \boldsymbol{S}^{-1}_t}{\partial \theta_j} \boldsymbol{C}_t \boldsymbol{m}^-_t + (\boldsymbol{m}^-_t)^T \boldsymbol{C}_t^T \boldsymbol{S}^{-1}_t \dfrac{\partial \boldsymbol{C}_t}{\partial \theta_j} \boldsymbol{m}^-_t \nonumber \\
&{+}\: (\boldsymbol{m}^-_t)^T \boldsymbol{C}_t^T \boldsymbol{S}^{-1}_t \boldsymbol{C}_t \dfrac{\partial \boldsymbol{m}^-_t}{\partial \theta_j}.
\end{align}
The second term in \eqref{eq:NLML_KF_derivative_app} can be calculated as
\begin{align}
  \dfrac{\partial }{\partial \theta_j} \log (\det(\boldsymbol{S}_t)) = \mathrm{tr} \left( \boldsymbol{S}_t^{-1} \dfrac{\partial \boldsymbol{S}_t}{\partial \theta_j} \right),
\end{align}
while the partial derivative of $\boldsymbol{S}_t^{-1}$ can be calculated as
\begin{align}
  \dfrac{\partial \boldsymbol{S}_t^{-1}}{\partial \theta_j} = -\boldsymbol{S}_t^{-1} \dfrac{\partial \boldsymbol{S}_t}{\partial \theta_j} \boldsymbol{S}_t^{-1},
\end{align}
see \cite{williams2006gaussian} for both identities.

We still need the partial derivatives of $\boldsymbol{S}_t$, $\boldsymbol{C}_t$, and $\boldsymbol{m}^-_t$.
Starting with the prior mean we have
\begin{align}
  \dfrac{\partial \boldsymbol{m}^-_t}{\partial \theta_j} {=}&\: \dfrac{\partial}{\partial \theta_j} \boldsymbol{A}_t \underbrace{\begin{tiny} \begin{pmatrix} \boldsymbol{f}^b_t \\
  \boldsymbol{m}_{t-1} + \boldsymbol{P}_{t-1} \left( \boldsymbol{A}^{f^b}_t \right)^T \left( \boldsymbol{S}^{f^b}_t \right)^{-1} \left( \boldsymbol{f}^b_t - \boldsymbol{A}^{f^b}_t \boldsymbol{m}_{t-1} \right) \end{pmatrix} \end{tiny}}_{=\tilde{\boldsymbol{m}}_{t-1}} \nonumber \\
  {=}&\: \dfrac{\partial \boldsymbol{A}_t}{\partial \theta_j} \tilde{\boldsymbol{m}}_{t-1} + \boldsymbol{A}_t \dfrac{\partial \tilde{\boldsymbol{m}}_{t-1}}{\partial \theta_j},
  \label{eq:priormeanderivatives}
\end{align}
with 
\begin{align} 
  &\: \dfrac{\partial \boldsymbol{A}_t}{\partial \theta_j} \nonumber \\
  \:{=}&\: \dfrac{\partial}{\partial \theta_j} \begin{bmatrix}\boldsymbol{K}^{ff^b}_{t,t} & \boldsymbol{K}^{ff}_{t,t-1} \end{bmatrix} 
  \begin{bmatrix} \boldsymbol{K}^{f^b f^b}_{t,t} & \boldsymbol{K}^{f^b f}_{t,t-1} \\
  \boldsymbol{K}^{ff^b}_{t-1,t} & \boldsymbol{K}^{ff}_{t-1,t-1} \end{bmatrix}^{-1} \nonumber \\
  \:{=}&\: \begin{bmatrix} \dfrac{\partial \boldsymbol{K}^{ff^b}_{t,t}}{\partial \theta_j} & \dfrac{\partial \boldsymbol{K}^{ff}_{t,t-1}}{\partial \theta_j} \end{bmatrix} 
  \begin{bmatrix} \boldsymbol{K}^{f^b f^b}_{t,t} & \boldsymbol{K}^{f^b f}_{t,t-1} \\
  \boldsymbol{K}^{ff^b}_{t-1,t} & \boldsymbol{K}^{ff}_{t-1,t-1} \end{bmatrix}^{-1} \nonumber \\
  &{+}\: \begin{bmatrix}\boldsymbol{K}^{f^b}_{t,t} & \boldsymbol{K}^{ff}_{t,t-1} \end{bmatrix} 
  \begin{bmatrix} \boldsymbol{K}^{f^b f^b}_{t,t} & \boldsymbol{K}^{f^b f}_{t,t-1} \\
  \boldsymbol{K}^{ff^b}_{t-1,t} & \boldsymbol{K}^{ff}_{t-1,t-1} \end{bmatrix}^{-1} \nonumber \\
  &{\times}\: \begin{bmatrix} \dfrac{\partial \boldsymbol{K}^{f^b f^b}_{t,t}}{\partial \theta_j} & \dfrac{\partial \boldsymbol{K}^{f^b f}_{t,t-1}}{\partial \theta_j} \\
  \dfrac{\partial \boldsymbol{K}^{ff^b}_{t-1,t}}{\theta_j} & \dfrac{\partial \boldsymbol{K}^{ff}_{t-1,t-1}}{\theta_j} \end{bmatrix}
  \begin{bmatrix} \boldsymbol{K}^{f^b f^b}_{t,t} & \boldsymbol{K}^{f^b f}_{t,t-1} \\
  \boldsymbol{K}^{ff^b}_{t-1,t} & \boldsymbol{K}^{ff}_{t-1,t-1} \end{bmatrix}^{-1},
\end{align}
\begin{align} 
\dfrac{\partial  \tilde{\boldsymbol{m}}_{t-1}}{\partial \theta_j} = \Bigg[ \boldsymbol{0},\: 
  \Bigg({}&\: \boldsymbol{P}_{t-1} \dfrac{\left( \boldsymbol{A}^{f^b}_t \right)^T}{\partial \theta_j} \left( \boldsymbol{S}^{f^b}_t \right)^{-1} \nonumber \\
 {}&{+}\: \boldsymbol{P}_{t-1} \left( \boldsymbol{A}^{f^b}_t \right)^T \dfrac{\partial \left( \boldsymbol{S}^{f^b}_t \right)^{-1}}{\partial \theta_j} \Bigg) \nonumber \\
  & \times \left( \boldsymbol{f}^b_t - \boldsymbol{A}^{f^b}_t \boldsymbol{m}_{t-1} \right) \nonumber \\
  & - \boldsymbol{P}_{t-1} \left( \boldsymbol{A}^{f^b}_t \right)^T \left( \boldsymbol{S}^{f^b}_t \right)^{-1} \dfrac{\partial \boldsymbol{A}^{f^b}_t}{\partial \theta_j} \boldsymbol{m}_{t-1} \Bigg]^T,
\end{align}
\begin{align} 
\dfrac{\partial \boldsymbol{A}^{f^b}_t}{\partial \theta_j} &= \dfrac{\partial}{\partial \theta_j} \left( \boldsymbol{K}^{f^b f}_{t,t-1} \left( \boldsymbol{K}^{ff}_{t-1,t-1} \right)^{-1} \right) \nonumber \\
&= \dfrac{\partial \boldsymbol{K}^{f^b f}_{t,t-1}}{\partial \theta_j} \left( \boldsymbol{K}^{ff}_{t-1,t-1} \right)^{-1} + \boldsymbol{K}^{f^b f}_{t,t-1} \dfrac{\partial \left( \boldsymbol{K}^{ff}_{t-1,t-1} \right)^{-1}}{\partial \theta_j},
\end{align}
\begin{align} 
\dfrac{\partial \boldsymbol{S}^{f^b}_t}{\partial \theta_j} \:{=}&\: \dfrac{\partial \boldsymbol{A}^{f^b}_t}{\partial \theta_j} \boldsymbol{P}_{t-1} \left( \boldsymbol{A}^{f^b}_t \right)^T + \dfrac{\partial \boldsymbol{P}_t^{GP,f^b}}{\partial \theta_j} \nonumber \\
&{+}\: \boldsymbol{A}^{f^b}_t \boldsymbol{P}_{t-1} \dfrac{\partial \left( \boldsymbol{A}^{f^b}_t \right)^T}{\partial \theta_j},
\end{align}
\begin{align} 
\dfrac{\partial \boldsymbol{P}^{GP,f^b}_t}{\partial \theta_j} &= \dfrac{\partial \boldsymbol{K}^{f^b f^b}_{t,t}}{\partial \theta_j} - \dfrac{\partial \boldsymbol{A}^{f^b}_t}{\partial \theta_j} \boldsymbol{K}^{ff^b}_{t-1,t} - \boldsymbol{A}^{f^b}_t \dfrac{\partial \boldsymbol{K}^{ff^b}_{t-1,t}}{\partial \theta_j}.
\end{align}

For the measurement matrix we have
\begin{align}
  \dfrac{\partial \boldsymbol{C}_t}{\partial \theta_j} \:{=}&\: \dfrac{\partial \boldsymbol{K}^{f^y f}_{tt} \left( \boldsymbol{K}^{ff}_{tt} \right)^{-1}}{\partial \theta_j} \nonumber \\
  \:{=}&\: \dfrac{\partial \boldsymbol{K}^{f^y f}_{tt}}{\partial \theta_j} \left( \boldsymbol{K}^{ff}_{tt} \right)^{-1} + \boldsymbol{K}^{f^y f}_{tt} \dfrac{\partial \left( \boldsymbol{K}^{ff}_{tt} \right)^{-1}}{\partial \theta_j} \nonumber \\
  \:{=}&\: \dfrac{\partial \boldsymbol{K}^{f^y f}_{tt}}{\partial \theta_j} \left( \boldsymbol{K}^{ff}_{tt} \right)^{-1} - \boldsymbol{K}^{f^y f}_{tt} \left( \boldsymbol{K}^{ff}_{tt} \right)^{-1} \dfrac{\partial \boldsymbol{K}^{ff}_{tt}}{\partial \theta_j} \left( \boldsymbol{K}^{ff}_{tt} \right)^{-1}.
\end{align}
The covariance matrix of the innovation is a bit more elaborate, we have
\begin{align}
  \dfrac{\partial \boldsymbol{S}_t}{\partial \theta_j} \:{=}&\: \dfrac{\partial \boldsymbol{C}_t \boldsymbol{P}^-_t \boldsymbol{C}_t^T + \boldsymbol{P}^{GP,f^y f^y}_t}{\partial \theta_j} \nonumber \\
  \:{=}&\: \dfrac{\partial \boldsymbol{C}_t}{\partial \theta_j} \boldsymbol{P}^-_t \boldsymbol{C}_t^T + \boldsymbol{C}_t \dfrac{\partial \boldsymbol{P}^-_t}{\partial \theta_j} \boldsymbol{C}_t^T \nonumber \\
  &{+}\: \boldsymbol{C}_t \boldsymbol{P}^-_t \dfrac{\partial \boldsymbol{C}_t^T}{\partial \theta_j} + \dfrac{\partial \boldsymbol{P}^{GP,f^y}_t}{\partial \theta_j}.
  \label{eq:innovationmatrixderivative}
\end{align}
The partial derivatives in the second and last term need to be derived.
For the prior covariance matrix of the state error we get
\begin{align}
  &\: \dfrac{\partial \boldsymbol{P}^-_t}{\partial \theta_j} \nonumber \\
  {=}&\: \dfrac{\partial \boldsymbol{A}_t}{\partial \theta_j} \tilde{\boldsymbol{P}}_{t-1} \boldsymbol{A}_t^T + \boldsymbol{A}_t \tilde{\boldsymbol{P}}_{t-1}  \dfrac{\partial \boldsymbol{A}_t^T}{\partial \theta_j} + \dfrac{\partial \boldsymbol{P}^{GP,f}_{t} }{\partial \theta_j} \nonumber \\
  &{-}\: \boldsymbol{A}_t \begin{bmatrix} \dfrac{\partial \boldsymbol{S}^{f^b}_t}{\partial \theta_j} \\ \boldsymbol{P}_{t-1} \dfrac{\partial \left( \boldsymbol{A}^{f^b}_t \right)^T}{\partial \theta_j} \end{bmatrix} \left( \boldsymbol{S}^{f^b}_t \right)^{-1} \begin{bmatrix} \boldsymbol{S}^{f^b}_t \\ \boldsymbol{P}_{t-1} \left( \boldsymbol{A}^{f^b}_t \right)^T \end{bmatrix}^T  \boldsymbol{A}_t^T \nonumber \\
  &{-}\: \boldsymbol{A}_t \begin{bmatrix} \boldsymbol{S}^{f^b}_t \\ \boldsymbol{P}_{t-1} \left( \boldsymbol{A}^{f^b}_t \right)^T \end{bmatrix} \dfrac{\partial \left( \boldsymbol{S}^{f^b}_t \right)^{-1}}{\partial \theta_j} \begin{bmatrix} \boldsymbol{S}^{f^b}_t \\ \boldsymbol{P}_{t-1} \left( \boldsymbol{A}^{f^b}_t \right)^T \end{bmatrix}^T  \boldsymbol{A}_t^T \nonumber \\
   &{-}\: \boldsymbol{A}_t \begin{bmatrix} \boldsymbol{S}^{f^b}_t \\ \boldsymbol{P}_{t-1} \left( \boldsymbol{A}^{f^b}_t \right)^T \end{bmatrix} \left( \boldsymbol{S}^{f^b}_t \right)^{-1} \begin{bmatrix} \dfrac{\partial \boldsymbol{S}^{f^b}_t}{\partial \theta_j} \\ \boldsymbol{P}_{t-1} \dfrac{\partial \left( \boldsymbol{A}^{f^b}_t \right)^T}{\partial \theta_j} \end{bmatrix}^T \boldsymbol{A}_t^T \nonumber \\
   &{-}\: \boldsymbol{A}_t \begin{bmatrix} \boldsymbol{S}^{f^b}_t \\ \boldsymbol{P}_{t-1} \left( \boldsymbol{A}^{f^b}_t \right)^T \end{bmatrix} \left( \boldsymbol{S}^{f^b}_t \right)^{-1} \begin{bmatrix} \boldsymbol{S}^{f^b}_t \\ \boldsymbol{P}_{t-1} \left( \boldsymbol{A}^{f^b}_t \right)^T \end{bmatrix}^T \dfrac{\partial \boldsymbol{A}_t^T}{\partial \theta_j}.
\end{align}
The partial derivatives of the dynamic matrix are already covered in \eqref{eq:priormeanderivatives}.
The partial derivative of the inherent GP uncertainty is
\begin{align}
  &\:\dfrac{\partial \boldsymbol{P}^{GP,f}_{t} }{\partial \theta_j} \nonumber \\
  {=}&\: \dfrac{\partial \boldsymbol{K}^{ff}_{t,t} - \boldsymbol{K}^{ff}_{t,t-1} \left( \boldsymbol{K}^{ff}_{t-1,t-1 } \right)^{-1}  \boldsymbol{K}^{ff}_{t-1,t}}{\partial \theta_j} \nonumber \\
  {=}&\: \dfrac{\partial \boldsymbol{K}^{ff}_{t,t}}{\partial \theta_j} - \dfrac{\partial \boldsymbol{K}^{ff}_{t,t-1}}{\partial \theta_j} \left( \boldsymbol{K}^{ff}_{t-1,t-1 } \right)^{-1}  \boldsymbol{K}^{ff}_{t-1,t} \nonumber \\
  &{+}\: \boldsymbol{K}^{ff}_{t,t-1} \left( \boldsymbol{K}^{ff}_{t-1,t-1} \right)^{-1} \dfrac{\partial \boldsymbol{K}^{ff}_{t-1,t-1} }{\partial \theta_j} \left( \boldsymbol{K}^{ff}_{t-1,t-1 } \right)^{-1}  \boldsymbol{K}^{ff}_{t-1,t} \nonumber \\
  &{-}\: \boldsymbol{K}^{ff}_{t,t-1} \left( \boldsymbol{K}^{ff}_{t-1,t-1 } \right)^{-1} \dfrac{\partial \boldsymbol{K}^{ff}_{t-1,t}}{\partial \theta_j}.  
\end{align} 
The partial derivative of the inherent GP uncertainty for the output values (last term of \ref{eq:innovationmatrixderivative}) is
\begin{align}
  \dfrac{\partial \boldsymbol{P}^{GP, f^y}_{t}}{\partial \theta_j} &= \dfrac{\partial \boldsymbol{K}^{f^y f^y}_{t,t} - \boldsymbol{K}^{f^y f}_{t,t} \left( \boldsymbol{K}^{f}_{t,t} \right)^{-1} \boldsymbol{K}^{ff^y}_{t,t}}{\partial \theta_j} \nonumber \\
  &= \dfrac{\partial \boldsymbol{K}^{f^y f^y}_{t,t}}{\partial \theta_j} - \dfrac{\partial \boldsymbol{K}^{f^y f}_{t,t}}{\partial \theta_j} \left( \boldsymbol{K}^{ff}_{t,t} \right)^{-1} \boldsymbol{K}^{ff^y}_{t,t} \nonumber \\
  &+ \boldsymbol{K}^{f^y f}_{t,t} \left( \boldsymbol{K}^{ff}_{t,t} \right)^{-1} \dfrac{\partial \boldsymbol{K}^{ff}_{t,t}}{\partial \theta_j} \left( \boldsymbol{K}^{ff}_{t,t} \right)^{-1} \boldsymbol{K}^{ff^y}_{t,t} \nonumber \\
  &- \boldsymbol{K}^{f^y f}_{t,t} \left( \boldsymbol{K}^{ff}_{t,t} \right)^{-1}  \dfrac{\partial \boldsymbol{K}^{ff^y}_{t,t}}{\partial \theta_j}.
\end{align}

\bibliographystyle{IEEEtran}
\bibliography{IEEEabrv,numericalGPKF-formal_bibs}

\begin{thebibliography}{10}
\providecommand{\url}[1]{#1}
\csname url@samestyle\endcsname
\providecommand{\newblock}{\relax}
\providecommand{\bibinfo}[2]{#2}
\providecommand{\BIBentrySTDinterwordspacing}{\spaceskip=0pt\relax}
\providecommand{\BIBentryALTinterwordstretchfactor}{4}
\providecommand{\BIBentryALTinterwordspacing}{\spaceskip=\fontdimen2\font plus
\BIBentryALTinterwordstretchfactor\fontdimen3\font minus
  \fontdimen4\font\relax}
\providecommand{\BIBforeignlanguage}[2]{{%
\expandafter\ifx\csname l@#1\endcsname\relax
\typeout{** WARNING: IEEEtran.bst: No hyphenation pattern has been}%
\typeout{** loaded for the language `#1'. Using the pattern for}%
\typeout{** the default language instead.}%
\else
\language=\csname l@#1\endcsname
\fi
#2}}
\providecommand{\BIBdecl}{\relax}
\BIBdecl

\bibitem{kalman1960new}
\BIBentryALTinterwordspacing
R.~E. Kalman, ``{A New Approach to Linear Filtering and Prediction Problems},''
  \emph{Journal of Basic Engineering}, vol.~82, no.~1, pp. 35--45, 03 1960.
  [Online]. Available: \url{https://doi.org/10.1115/1.3662552}
\BIBentrySTDinterwordspacing

\bibitem{curtain2012introduction}
R.~F. Curtain and H.~Zwart, \emph{An introduction to infinite-dimensional
  linear systems theory}.\hskip 1em plus 0.5em minus 0.4em\relax Springer
  Science \& Business Media, 2012, vol.~21.

\bibitem{ferziger2002computational}
J.~H. Ferziger, M.~Peri{\'c}, and R.~L. Street, \emph{Computational methods for
  fluid dynamics}.\hskip 1em plus 0.5em minus 0.4em\relax Springer, 2002,
  vol.~3.

\bibitem{raissi2018numerical}
M.~Raissi, P.~Perdikaris, and G.~E. Karniadakis, ``{Numerical Gaussian
  processes for time-dependent and nonlinear partial differential equations},''
  \emph{SIAM Journal on Scientific Computing}, vol.~40, no.~1, pp. A172--A198,
  2018.

\bibitem{kueper2020mtns}
A.~K{\"u}per, N.~Totis, and S.~Waldherr, ``{Comparing cell population balance
  model simulation through Gaussian processes and discretisation},'' in
  \emph{Proceedings of the 24th International Symposium on Mathematical Theory
  of Networks and Systems [not yet published]}.\hskip 1em plus 0.5em minus
  0.4em\relax IFAC, 2020.

\bibitem{williams2006gaussian}
C.~K. Williams and C.~E. Rasmussen, \emph{Gaussian processes for machine
  learning}.\hskip 1em plus 0.5em minus 0.4em\relax MIT Press Cambridge, MA,
  2006, vol.~2, no.~3.

\bibitem{kueper2020ifacnGPKF}
\BIBentryALTinterwordspacing
A.~Küper and S.~Waldherr, ``{Numerical Gaussian process Kalman filtering},''
  \emph{IFAC-PapersOnLine}, vol.~53, no.~2, pp. 11\,416--11\,421, 2020, 21th
  IFAC World Congress. [Online]. Available:
  \url{https://www.sciencedirect.com/science/article/pii/S2405896320308788}
\BIBentrySTDinterwordspacing

\bibitem{sarkka2012infinite}
S.~S{\"a}rkk{\"a} and J.~Hartikainen, ``{Infinite-dimensional Kalman filtering
  approach to spatio-temporal Gaussian process regression},'' in
  \emph{International Conference on Artificial Intelligence and Statistics},
  2012, pp. 993--1001.

\bibitem{sarkka2013spatiotemporal}
S.~S{\"a}rkk{\"a}, A.~Solin, and J.~Hartikainen, ``{Spatiotemporal learning via
  infinite-dimensional Bayesian filtering and smoothing: A look at Gaussian
  process regression through Kalman filtering},'' \emph{IEEE Signal Processing
  Magazine}, vol.~30, no.~4, pp. 51--61, 2013.

\bibitem{TODESCATO2020109032}
M.~Todescato, A.~Carron, R.~Carli, G.~Pillonetto, and L.~Schenato, ``{Efficient
  spatio-temporal Gaussian regression via Kalman filtering},''
  \emph{Automatica}, vol. 118, p. 109032, 2020.

\bibitem{ko2009gp}
J.~Ko and D.~Fox, ``{GP-BayesFilters: Bayesian filtering using Gaussian process
  prediction and observation models},'' \emph{Autonomous Robots}, vol.~27,
  no.~1, pp. 75--90, 2009.

\bibitem{deisenroth2009analytic}
M.~P. Deisenroth, M.~F. Huber, and U.~D. Hanebeck, ``{Analytic moment-based
  Gaussian process filtering},'' in \emph{Proceedings of the 26th annual
  international conference on machine learning}.\hskip 1em plus 0.5em minus
  0.4em\relax ACM, 2009, pp. 225--232.

\bibitem{deisenroth2011general}
M.~P. Deisenroth and H.~Ohlsson, ``{A general perspective on Gaussian filtering
  and smoothing: Explaining current and deriving new algorithms},'' in
  \emph{Proceedings of the 2011 American Control Conference}.\hskip 1em plus
  0.5em minus 0.4em\relax IEEE, 2011, pp. 1807--1812.

\bibitem{sarkka2011linear}
S.~S{\"a}rkk{\"a}, ``{Linear operators and stochastic partial differential
  equations in Gaussian process regression},'' in \emph{International
  Conference on Artificial Neural Networks}.\hskip 1em plus 0.5em minus
  0.4em\relax Springer, 2011, pp. 151--158.

\bibitem{sarkka2019applied}
S.~S{\"a}rkk{\"a} and A.~Solin, \emph{Applied stochastic differential
  equations}.\hskip 1em plus 0.5em minus 0.4em\relax Cambridge University
  Press, 2019, vol.~10.

\bibitem{curtain1975survey}
R.~Curtain, ``A survey of infinite-dimensional filtering,'' \emph{Siam Review},
  vol.~17, no.~3, pp. 395--411, 1975.

\bibitem{simon2006optimal}
D.~Simon, \emph{{Optimal state estimation: Kalman, H infinity, and nonlinear
  approaches}}.\hskip 1em plus 0.5em minus 0.4em\relax John Wiley \& Sons,
  2006.

\bibitem{sarkka2013bayesian}
S.~S{\"a}rkk{\"a}, \emph{Bayesian filtering and smoothing}.\hskip 1em plus
  0.5em minus 0.4em\relax Cambridge University Press, 2013, vol.~3.

\bibitem{fitzgerald1971divergence}
R.~Fitzgerald, ``{Divergence of the Kalman filter},'' \emph{IEEE Transactions
  on Automatic Control}, vol.~16, no.~6, pp. 736--747, 1971.

\bibitem{zeng2019sample}
S.~Zeng, ``Sample-based population observers,'' \emph{Automatica}, vol. 101,
  pp. 166--174, 2019.

\bibitem{zeng2015ensemble}
S.~Zeng, S.~Waldherr, C.~Ebenbauer, and F.~Allg{\"o}wer, ``Ensemble
  observability of linear systems,'' \emph{IEEE Transactions on Automatic
  Control}, vol.~61, no.~6, pp. 1452--1465, 2015.

\bibitem{purisha2019probabilistic}
Z.~Purisha, C.~Jidling, N.~Wahlstr{\"o}m, T.~B. Sch{\"o}n, and
  S.~S{\"a}rkk{\"a}, ``Probabilistic approach to limited-data computed
  tomography reconstruction,'' \emph{Inverse Problems}, vol.~35, no.~10, p.
  105004, 2019.

\bibitem{snelson2006sparse}
E.~Snelson and Z.~Ghahramani, ``{Sparse Gaussian processes using
  pseudo-inputs},'' \emph{Advances in Neural Information Processing Systems},
  vol.~18, pp. 1259--1266, 2006.

\bibitem{solin2020hilbert}
A.~Solin and S.~S{\"a}rkk{\"a}, ``{Hilbert space methods for reduced-rank
  Gaussian process regression},'' \emph{Statistics and Computing}, vol.~30,
  no.~2, pp. 419--446, 2020.

\end{thebibliography}

\end{document}